\documentclass[fleqn,usenatbib]{mnras}
\usepackage[T1]{fontenc}

\usepackage{graphicx}
\DeclareGraphicsExtensions{.pdf}
\usepackage{amsmath}
\usepackage{amssymb}
\usepackage{tabularx}

\usepackage{hyperref}
\hypersetup{
    colorlinks=true,
    linkcolor=red,
    citecolor=blue,
    urlcolor=blue,
}
\usepackage{tablefootnote}

\usepackage{booktabs} 
\usepackage{subcaption}
\usepackage{caption}
\usepackage[english=american]{csquotes}

\newcommand{\cmfast}{\textsc{{\tt 21cmFAST}}}
\newcommand\lsim{\mathrel{\rlap{\lower4pt\hbox{\hskip1pt$\sim$}}
        \raise1pt\hbox{$<$}}}
\newcommand\gsim{\mathrel{\rlap{\lower4pt\hbox{\hskip1pt$\sim$}}
        \raise1pt\hbox{$>$}}}
\newcommand{\Rom}[1]{\uppercase\expandafter{\romannumeral #1}}
\newcommand{\rom}[1]{\lowercase\expandafter{\romannumeral #1}}

\renewcommand{\b}{\mathbf}

\usepackage[dvipsnames]{xcolor}
\definecolor{GREEN}{rgb}{0.0, 0.7, 0.0}

\definecolor{DARKORANGE}{rgb}{1.0, 0.5490196078431373, 0.0}
\definecolor{ROYALBLUE}{rgb}{0.2549019607843137, 0.4117647058823529, 0.8823529411764706}
\definecolor{INDIGO}{rgb}{0.29411764705882354, 0.0, 0.5098039215686274}
\newcommand{\cCNN}[1]{{\color{DARKORANGE} #1}}
\newcommand{\cSRNN}[1]{{\color{ROYALBLUE} #1}}
\newcommand{\cCRNN}[1]{{\color{INDIGO} #1}}

\title[RNNs for 21 cm lightcones]{Machine learning  astrophysics from 21 cm lightcones: impact of network architectures and signal contamination}
\author[Prelogović et al.]{David Prelogović$^{1}$\thanks{E-mail: david.prelogovic@sns.it}, Andrei Mesinger$^1$, Steven Murray$^2$, Giuseppe Fiameni$^{3}$, and \newauthor{Nicolas Gillet$^4$}\\
	$^{1}$ Scuola Normale Superiore, Piazza dei Cavalieri 7, 56126 Pisa, Italy\\
	$^{2}$ School of Earth and Space Exploration, Arizona State University, 781 Terrace Mall, Tempe, AZ, 85287, USA \\
	$^{3}$ NVIDIA AI Technology Center Italy, Via Melchiorre Gioia 8, 820124 Milan, Italy\\
	$^{4}$ Observatoire Astronomique de Strasbourg, Université de Strasbourg, 11 rue de l’Université, 67000 Strasbourg, France
}
\date{Accepted XXX. Received YYY; in original form ZZZ}
\pubyear{2021}

\begin{document}

\label{firstpage}
\pagerange{\pageref{firstpage}--\pageref{lastpage}}
\maketitle

\begin{abstract}
 Imaging the cosmic 21 cm signal will map out the first billion years of our Universe. The resulting 3D lightcone (LC) will encode the properties of the unseen first galaxies and physical cosmology. 
Here we build on previous work using Neural Networks (NNs) to infer astrophysical parameters directly from 21 cm LC images.  We introduce Recurrent Neural Networks (RNNs), capable of efficiently characterizing the evolution along the redshift axis of 21 cm LC images.  Using a large database of simulated cosmic 21 cm LCs, we compare the relative performance in parameter estimation of different network architectures.  These including two types of RNNs, which differ in their complexity, as well as a more traditional Convolutional Neural Network (CNN).
For the ideal case of no instrumental effects, our simplest and easiest to train RNN performs the best, with a mean squared parameter estimation error (MSE) that is lower by a factor of $\gsim$ 2 compared with the other architectures studied here, and a factor of $\gsim$ 8 lower than the previously-studied CNN.
We also corrupt the cosmic signal by adding noise expected from a 1000\,h integration with the Square Kilometre Array, as well as excising a foreground-contaminated \enquote{horizon wedge}.
Parameter prediction errors increase when the NNs are trained on these contaminated LC images, though recovery is still good even in the most pessimistic case (with $R^2 \gsim 0.5 - 0.95$).  However, we find no notable differences in performance between network architectures on the contaminated images.
We argue this is due to the size of our dataset, highlighting the need for larger datasets and/or better data augmentation in order to maximize the potential of NNs in 21 cm parameter estimation.

\end{abstract}


\section{Introduction}

Tomography using the hyperfine transition of \textsc{Hi} is set to revolutionize our studies of the Cosmic Dawn (CD) and subsequent Epoch of Reionization (EoR).  Although current radio telescopes are aiming for a statistical detection, the upcoming Square Kilometre Array (SKA)\footnote{\url{https://www.skatelescope.org}} will eventually provide a 3D image of the first billion years of our Universe.  This image encodes the properties of the first generations of galaxies, whose UV and X-ray radiation fields imprint multi-scale patterns in the 21 cm signal (see a recent review in \citealt{Mesinger2019}).

How can we best interpret these patterns to learn about astrophysics and cosmology? 
Bayesian inference has recently become established in the field, either by directly forward-modeling the 21 cm lightcone (e.g. \citealt{greig17, greig18, park19, Greig21}) or through the use of emulators of the 21 cm power spectrum (e.g. \citealt{Kern17, SP18, jennings19, Ghara20, Mondal20}). However, the question of most constraining summary statistic to use when comparing theory to data is as-yet unsettled.   The huge data volumes and the fact that we do not know the initial seed of the Universe, necessitate some form of compression of 21 cm images. However, the fact that the images are notably non-Gaussian (e.g. see Fig.~1 in \citealt{mellema15}) means that there is additional information contained in the phases of wave-modes that is ignored in the commonly-used power spectrum statistic. Indeed several studies have shown that non-Gaussian and morphological statistics contain complimentary information, and can improve parameter inference when combined with the power spectrum (e.g. \citealt{gazagnes21, watkinson21}).

So what is the \enquote{optimal} statistic for constraining astrophysics and cosmology from 21 cm images? Several candidate statistics (e.g. \citealt{shimabukuro2015studying,shimabukuro2017constraining,majumdar2018quantifying,giri2018a,giri2019,gorce2019studying,watkinson2019,watkinson21,gazagnes21}) have been investigated in the literature, but these investigations have not performed a systematic treatment of the \enquote{optimality} of the various statistics. 
Indeed without a strong a priori physical motivation, an optimal statistics is unlikely to be found.

An alternative approach is provided by deep learning techniques, specifically neural networks (NN). By minimizing a loss function (the prediction error) NN can adaptively find a summary statistic that provides the most accurate parameter recovery. This comes with the downside that the resulting compression is difficult to interpret physically, and thus the performance strongly depends on having a large, representative database for training.

 NNs are rapidly becoming popular in the 21 cm community, recovering the underlying astrophysics and cosmology from 21 cm images.  \citet{gillet19} introduced NN to the field, using 2D Convolutional Neural Networks (CNNs, \citealt{lecun1989backpropagation}) on idealized lightcone images.  Although they did not consider instrumental effects, the database of \citet{gillet19} spanned the largest variation in the cosmic signal, as they considered both EoR and X-ray heating parameters.
\citealt{21cm_ML_LaPlante19} used a similar 2D CNN architecture to predict EoR timing and duration after removing foreground-contaminated modes. \citealt{21cm_ML_Kwon20} predicted the mean neutral fraction during the EoR from 2D slices contaminated with Gaussian SKA noise. Better realism of SKA1-Low telescope effects was then used with 2D CNNs to constrain cosmology and astrophysics \citep{hassan2020} and the neutral fraction \citep{21cm_ML_Magena20}. Prediction uncertainties were introduced by \citealt{HMV20}, using Bayesian Neural Networks to constrain cosmological and reionization parameters, however without instrumental effects. \citealt{zhao2021} on the other hand used likelihood free Bayesian inference to retrieve posteriors of two idealized reionization parameters. 
In addition to these regression studies, CNN-based UNets were used to identify ionized regions during the EoR and remove foreground contamination from mock 21 cm images (e.g. \citealt{21cm_ML_Makinen20, bianco2021, Gargon-Hartman21}).

However, the importance of the network architecture is often under-appreciated. The above studies use CNNs to process 21 cm images, since they are capable of picking up correlations over a range of scales.  Indeed CNNs are very popular in computer vision, such as object detection, image semantic segmentation and facial recognition.  However, EoR/CD radio images have rather unusual properties: they are three dimensional and {\it anisotropic}.  Foreground and instrument systematics leave different imprints in the sky-plane and the frequency axis.  More fundamentally, the lightcones of the cosmic signal have redshift evolution along the frequency axis, introducing strong correlations between neighboring frequency bins.

In this work, we introduce long short term memory (LSTM, \citealt{LSTM}) recurrent neural networks (RNNs) for parameter estimation from 21 cm images.  RNNs efficiently encode local correlations in sequential data, that would otherwise require deeper (and thus more complex) network layers (for details see Appendix \ref{app:layers}).  As such, RNNs have become very popular in applications with temporal evolution between images/frames (e.g. language, audio, video; see the review in \citealt{Schmidt19}). 
Here we exploit the directional differences in 21 cm lightcones by coupling ``traditional'' 2D CNNs for the sky-plane with recurrent layers for the frequency dimension that encodes space-time evolution.
We compare the performance of different architectures, including RNNs and  a variant of 3D CNNs, in parameter estimation from 21 cm lightcones with varying degrees of signal contamination: (i) the cosmic signal only; (ii) signal + telescope noise; (iii) signal + telescope noise + foreground wedge excision. 
All architectures and relevant functions used in this work are publicly available in \texttt{21cmRNN}\footnote{\url{https://github.com/dprelogo/21cmRNN}} package. 

This paper is organized as follows. In \S\ref{ch:mock_21cm_images} we discuss how 21 cm datasets are generated. The motivation behind different network architectures is discussed in \S\ref{ch:network_architectures}. The training procedure and performance are presented in \S\ref{sec:training}. In \S\ref{sec:parameter_recovery} we quantify the parameter recovery for different architectures and levels of signal contamination.  Finally, concluding remarks are presented in \S\ref{ch:discussion_and_conclusions}. Additional details on LSTM structure and architectures used can be found in Appendices \S\ref{app:layers} \& \S\ref{app:DetailedArchitectures}, respectively. All quantities are quoted in co-moving units assuming $\Lambda\mathrm{CDM}$ cosmology: $\left(\Omega_{\Lambda}, \Omega_{\mathrm{M}}, \Omega_{b}, n, \sigma_{8}, H_{0}\right) =\left(0.69,0.31,0.048,0.97,0.81,68 \mathrm{~km} \mathrm{~s}^{-1} \mathrm{Mpc}^{-1}\right)$, consistent with the results from \cite{planck_2018}. 

\section{Databases of mock 21 cm images} \label{ch:mock_21cm_images}

\begin{figure}
    \centering
    \includegraphics[width=\linewidth]{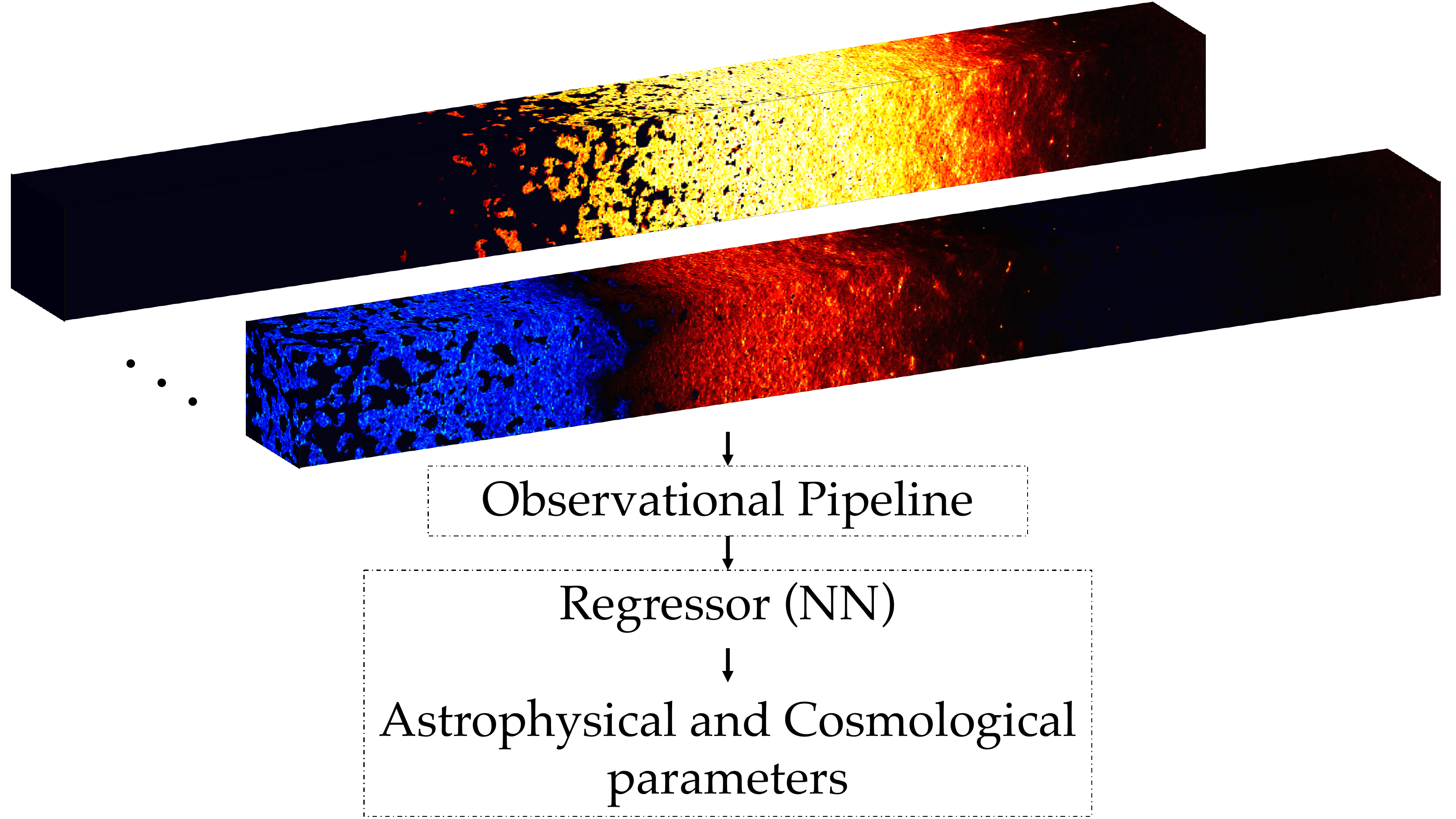}
    \caption{Generic pipeline for parameter estimation with NNs. A database of cosmic signal lightcones is processed through an observational pipeline, including noise and systematics. The resulting database of 21 cm images is fed into a regressor (neural network), trained to predict a “best guess” for astrophysical and cosmological parameters.}
    \label{fig:processing}
\end{figure}

Figure \ref{fig:processing} shows the sketch of our procedure for creating a database of mock 21 cm lightcones.  We sample astrophysical parameters and cosmological initial seeds to compute $3\mathrm{D}$ lightcones of the cosmological 21 cm signal (\S\ref{sec:cosmo}).  Each lightcone is then passed through an observational pipeline (\S\ref{sec:obs}), including the following steps: (i) mean removal; (ii) addition of thermal noise; (iii) removal of the foreground wedge.  As a result, we generate three databases, corresponding to steps (i), (i)+(ii) and (i)+(ii)+(iii).  These are then used to train a NN to predict astrophysical parameters.  Below we describe each of these steps in turn.  

\subsection{Cosmological 21 cm signal} \label{sec:cosmo}

The cosmological 21 cm signal is defined as the brightness temperature offset with respect to the CMB, $\delta T_b \equiv T_b - T_\gamma$. It can be expressed as (e.g. \citealt{Furlanetto06}):
\begin{equation}
    \begin{aligned}
        \delta T_{b}(\nu) &= \frac{T_{\mathrm{S}}-T_{\gamma}}{1+z}\left(1-e^{-\tau_{\nu_{0}}}\right) \\
        &\approx 27 \, x_{\mathrm{HI}}\left(1+\delta_{\mathrm{nl}}\right)\left(\frac{H}{\mathrm{d} v_{r} / \mathrm{d} r+H}\right)\left(1-\frac{T_{\gamma}}{T_{\mathrm{S}}}\right) \times \\
        & \times\left(\frac{1+z}{10} \frac{0.15}{\Omega_{\mathrm{M}} h^{2}}\right)^{1 / 2}\left(\frac{\Omega_{b} h^{2}}{0.023}\right)\, \mathrm{[mK]} \, .
    \end{aligned}
    \label{eq:21 cm_signal}
\end{equation}
Here, $T_{\mathrm{S}}$ is the gas spin temperature, defining the 21 cm level population as $n_1 / n_0 = 3 \exp(-0.068\mathrm{K} / T_{\mathrm{S}})$, $\tau_{\nu_{0}}$ is the optical depth at the 21 cm frequency $\nu_{0}$, $1 + \delta_{\mathrm{nl}}(\mathbf{x}, z) \equiv \rho / \bar{\rho}$ is the overdensity, $H(z)$ the Hubble parameter and $\mathrm{d} v_{r} / \mathrm{d} r$ the comoving gradient of the line of sight component of the comoving velocity. All quantities are evaluated at redshift $z=\nu_{0} / \nu-1$.

We use the database of cosmic 21 cm signals from \cite{gillet19}, consisting of $10,000$ lightcone images with a box size of 300\,Mpc and spatial resolution of 1.5\,Mpc, simulated using the public, semi-numerical code \texttt{\cmfast{} v2}\footnote{\url{https://github.com/andreimesinger/21 cmFAST}} \citep{21cmFAST_Mesinger_07, 21cmFAST_Mesinger11}.  
For a given set of astrophysical parameters and choice of cosmological initial seed, the code generates a 3D lightcone of $\delta T_{b}(x, y, \nu)$, where the first two dimensions correspond to on-sky coordinates, while the third corresponds to the frequency (redshift) dimension (see Fig. \ref{fig:processing}).  This calculation involves generating the initial density and velocity fields, which are then evolved using Lagrangian Perturbation Theory (e.g. \citealt{zeldovich1970}).  The spatially-dependent source field is computed from the evolved density using conditional halo mass functions (e.g. \citealt{barkana2004}).  For a given set of galaxy parameters, the inhomogenous reionization field is obtained by comparing the cumulative number of ionizing photons to the number of recombinations (e.g. \citealt{sobacchi2014}) in regions of decreasing radii (e.g. \citealt{furlanetto2004}).  Photons with long mean free paths, such as the soft UV and X-rays, are instead tracked by integrating the local emissivity back along the lightcone, for each simulation cell.  Soft UV photons are attenuated using "picket-fence" IGM absorption by the Lyman lines, while X-ray photons are attenuated by the partially ionized hydrogen and helium in the neutral component of the two-phased IGM.  For more details on these calculations, interested readers are encouraged to see \citep{21cmFAST_Mesinger_07, 21cmFAST_Mesinger11}.

The database of \cite{gillet19} varies four astrophysical parameters (in addition to co-varying the random seed), chosen to have both a clear physical meaning as well as driving the largest expected variation in the signal:
\begin{itemize}
    \item $\zeta \in [10, 250] \, ,$ the UV ionizing efficiency of galaxies. It mainly controls the timing of the EoR, where higher values ionize the Universe earlier. It can be expanded as
    \begin{equation}
        \zeta = 30 \, \frac{f_{\mathrm{esc}}}{0.1} \, \frac{f_*}{0.05} \, \frac{N_{\gamma/\mathrm{b}}}{4000} \, \frac{1.5}{1 + n_{\mathrm{rec}}} \, ,
    \end{equation}
    where the RHS corresponds to the following population-averaged quantities: $f_{\mathrm{esc}}$ is the fraction of ionizing photons escaping the host galaxy into the IGM, $f_*$ is the fraction of galactic gas in stars, $N_{\gamma/\mathrm{b}}$ number of photons per baryon produced in stars and $n_{\mathrm{rec}}$ is the average number of times a hydrogen atom recombines during the EoR.  We use a constant $\zeta$ in order to allow direct comparison with previous work, though we note that \texttt{\cmfast{} v2} allows $\zeta$ to scale with halo mass.

    \item $T_{\mathrm{vir}} \in [10^4, 10^6]\, \mathrm{K} \, ,$ the minimum virial temperature of halos hosting efficiently star forming galaxies.  Smaller halos have suppressed star formation due to inefficient cooling and/or feedback.  $T_\mathrm{vir}$ controls timing of all of astrophysical epochs.  Moreover, it also impacts the characteristic scales of the heated/ionized regions, since it parametrizes the typical bias of the relevant galaxy population.
  
    \item $L_{\mathrm{X<2keV}}/\mathrm{SFR} \in [10^{38}, 10^{42}]\, \mathrm{erg \, s^{-1} \, M_{\odot}^{-1} \, yr\, ,}$ the soft-band (with energies $<2$keV) X-ray luminosity per unit star formation rate (SFR). X-rays are responsible for heating the neutral IGM pre-reionization, and the resulting temperature fluctuations could drive the largest variance in the 21 cm signal (e.g. \citealt{Mesinger14}). $L_{\mathrm{X<2keV}}/\mathrm{SFR}$ controls the timing of the epoch of heating (EoH) and correspondingly the overlap between the other astrophysical epochs of EoR and Lyman alpha coupling.  Because of the cross terms in these fields, the level of overlap impacts the variance and dynamic range of the 21 cm signal.
    
    \item $E_0 \in [0.1, 1.5] \, \mathrm{keV} \, ,$ the minimum energy of X-ray photons escaping the host galaxy. Photons of lower energies are absorbed in the ISM of the host galaxies, and thus do not contribute to the EoH.  Because $E_0$ determines the typical mean free path of X-ray photons, it influences the timing and morphology of EoH, i.e. how homogeneously the IGM is heated. 
\end{itemize}
 Datasets are generated by sampling from flat distributions over the quoted ranges. $T_{\mathrm{vir}}$ and $L_{\mathrm{X<2keV}}/\mathrm{SFR}$ are sampled in log space, while $\zeta$ and $E_0$ are sampled in linear space.

We stress that it is important to model all of the astrophysical epochs for robust inference from 21 cm images.  In particular, it is common for  machine learning studies to focus only on the EoR, under the assumption that  $T_\mathrm{S} \gg T_\gamma$ is valid during the EoR (c.f. Eq. \ref{eq:21 cm_signal}; though see \citealt{gillet19, HMV20} who self-consistently compute the temperature).  However, recent studies have pointed out that $T_\mathrm{S} \gg T_\gamma$ is unlikely to be true during most of reionization.  The observed galaxy UV luminosity functions (LF) imply a decreasing star formation efficiency with halo mass (e.g. \citealt{mirocha17, park20}), that in turn suggests a later epoch of heating than initial estimates.  Although too simplistic to fully describe high-$z$ galaxies, the parametrization we use in this proof-of-concept study does contain ``tuning knobs'' for the relative timings and morphologies of all astrophysical epochs probed by the cosmic 21 cm signal.

\subsection{Simple observational pipeline} \label{sec:obs}
Starting from the database of cosmic 21 cm lightcones described in the previous section, we add instrumental effects with the following three steps:
\begin{itemize}
    \item {\it Mean removal} - remove the mean of the signal for each frequency slice,
    \item {\it Instrumental noise} - compute the $uv$ coverage and sample a realization of the instrumental noise for a $1000 \, \mathrm{h}$ measurement with SKA1-Low,
    \item {\it Wedge removal} - remove from the image all Fourier modes residing in a foreground-contaminated ``wedge''.
\end{itemize}

The first step is simply the result of measuring the signal with an interferometer, thus losing the  global signal (we label it as $\delta \widetilde{T}_b$). The last two steps are further detailed below.

\subsubsection{Instrumental Noise}
For our archetypal 21 cm interferometer we use SKA1-Low, whose design is optimized for high signal-to-noise (S/N) images \citep{dewdney2013ska1}.
The $uv$ coverage and instrumental noise are  calculated using \texttt{tools21cm}\footnote{\url{https://github.com/sambit-giri/tools21cm}} \citep{giri2018,giri2020}. We assume a tracked scan of $t_{\mathrm{daily}} = 6 \, \mathrm{h/day}$, $t_{\mathrm{int}} = 10 \, \mathrm{s}$ integration time, and a total $t_{\mathrm{obs}} = 1000 \, \mathrm{h}$ measurement.
For computational convenience, we fix the $uv$ grid to the Fourier dual grid of the lightcone and use a box-car as a gridding kernel.
With this approximation, the thermal noise is computed for each frequency slice based on the total time spent in each uv cell by all baselines throughout the $1000\,\mathrm{h}$ measurement (accounting for Earth's rotation):
\begin{equation}
\label{eq:noise}
    \sigma_{uv} = 
    \frac{T_{\rm sys}\cdot \Omega_{\rm beam} / \Omega_{\rm pix}}{\sqrt{2 \Delta \nu \,t_{\rm int}} }
    \cdot
    \frac{1}{\sqrt{N_{uv} \cdot t_{\mathrm{obs}} / t_{\mathrm{daily}}}} \,\,{\rm [mK]} \, ,
\end{equation}
where the first factor on the RHS corresponds to the noise over a Fourier-cell from a single baseline, expressed in temperature units (e.g. \citealt{parsons12}), and the second factor accounts for multiple measurements of a given $uv$ cell: $N_{uv}$ integration times per night for $ t_{\mathrm{obs}} / t_{\mathrm{daily}}$ nights. Thus, total observation time is discretized in visibility snapshots $t_\mathrm{int}$ apart.
The effective beam solid angle, $\Omega_{\rm beam} \sim \lambda^2 / A_{\rm eff} \approx 0.004 (\nu/150\, {\rm MHz})^{-2}\, {\rm sr}$ encodes the collecting area of the instrument, while the pixel solid angle accounts for the fact that the noise is inherently an angle-integrated quantity.
$T_{\rm sys}$ for the SKA is given by
\begin{equation}
    T_{\rm sys} = 60 \left(\frac{\nu}{300\,{\rm MHz}}\right)^{-2.55}\ \ [{\rm K}].
\end{equation}

A realization of the observed, gridded visibilities including thermal noise is computed as: $\delta\widetilde{T}_b (\b{u}, \nu)_\mathrm{obs} = \delta\widetilde{T}_b (\b{u}, \nu)_\mathrm{cosmo} + \mathcal{N}(\mu = 0, \sigma = \sigma_{uv})$, where $\delta\widetilde{T}_b (\b{u}, \nu)_\mathrm{cosmo}$ is the simulated brightness temperature Fourier-transformed in the sky-plane (using the Discrete Fourier Transform convention in which no length-normalization is applied), and $\mathcal{N}$ is a random variable drawn from a zero-mean Gaussian distribution with variance $\sigma^2_{uv}$.

\begin{figure*}
    \centering
    \includegraphics[width=\linewidth]{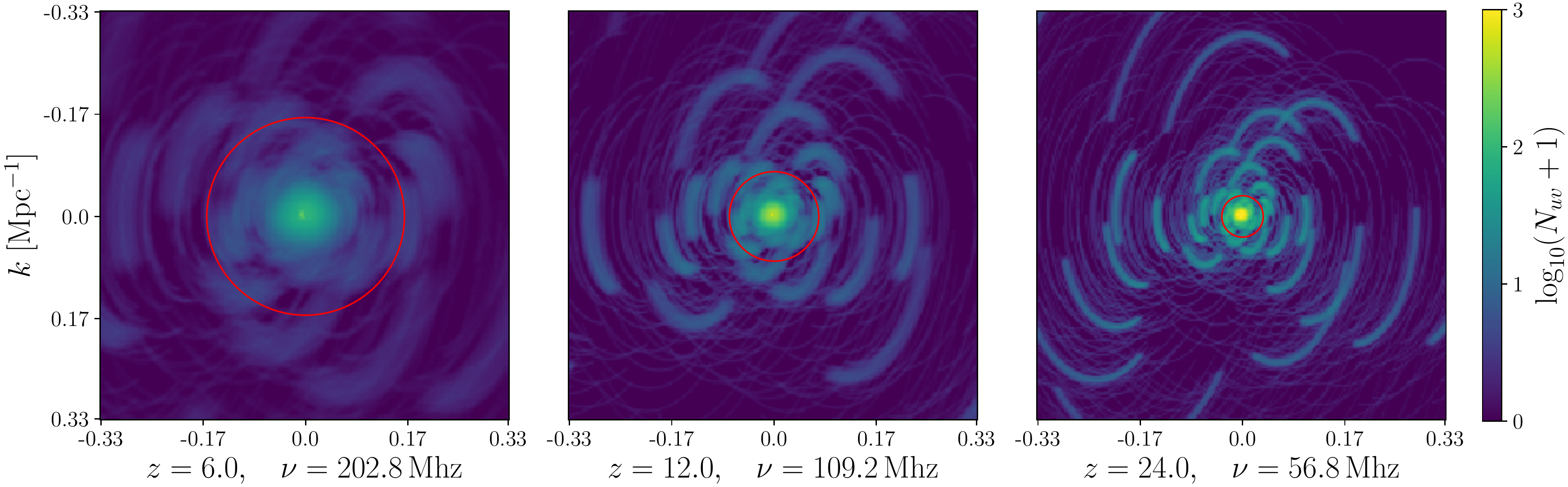}
    \caption{Daily $uv$ coverage at different redshifts (frequencies). The red circle denotes the maximum baselines considered when computing the noise ($2 \, \mathrm{km}$). $N_{uv}$ represents the number of  measurements in a given $uv$ cell, for $t_\mathrm{obs} = 6 \, \mathrm{h}$, $t_\mathrm{int} = 10 \, \mathrm{s}$.}
    \label{fig:uv_coverage}
\end{figure*}

In Figure \ref{fig:uv_coverage} we show the $uv$ coverage for a daily, $6\,\mathrm{h}$ observation, at different redshifts. When computing the thermal noise in Eq. \ref{eq:noise}, we use only the core stations: baselines shorter than $2 \, \mathrm{km}$, marked with a red circle, that provide most of the S/N.  We additionally only consider cells with $N_{uv} \geq 15$, roughly amounting to one full day of observation.  For simplicity, our frequency bins match the native resolution of the cosmological simulation, $\Delta\nu(z) = 1.5 \, \mathrm{Mpc}$.

\begin{figure}
    \centering
    \includegraphics[width=\linewidth]{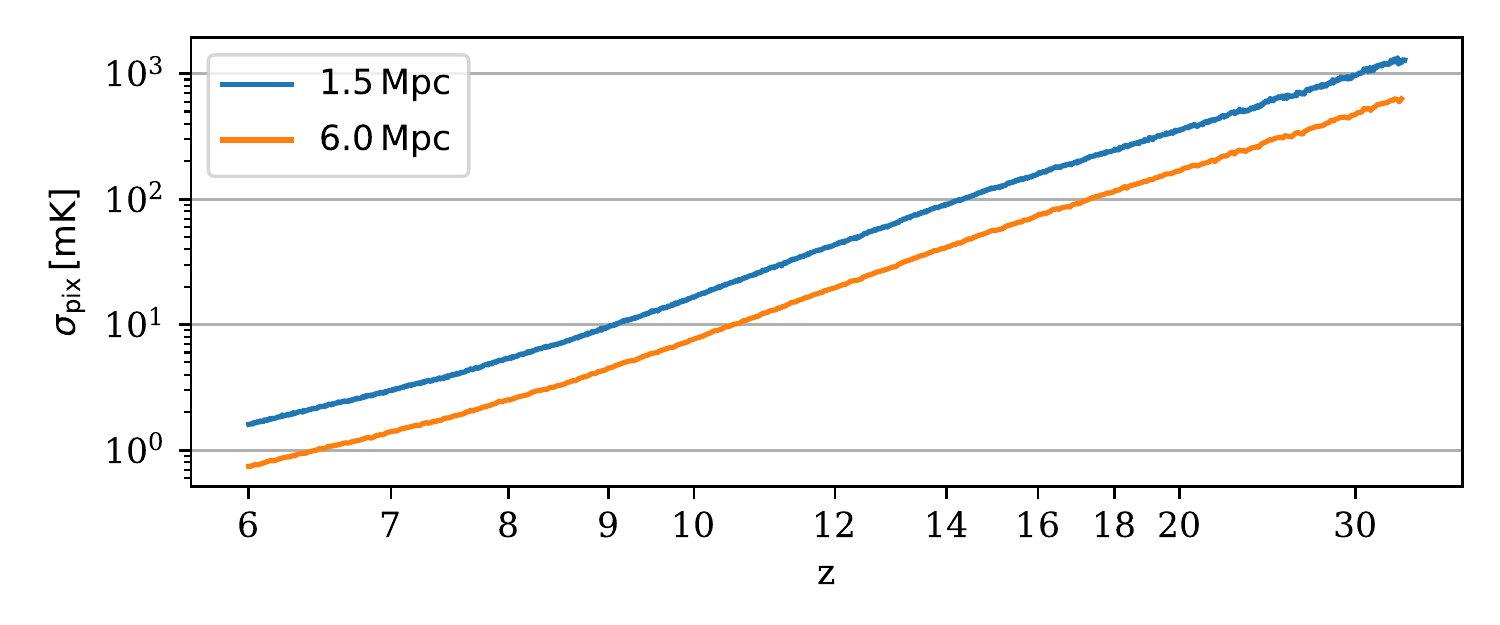}
    \caption{Telescope noise in $\mathrm{mK}$ at a pixel level as a function of redshift for our fiducial 1000h SKA observation. The two curves correspond to pixels of side-lengths $1.5$ and $6 \, \mathrm{Mpc}$.}
    \label{fig:rms_z}
\end{figure}

In Figure \ref{fig:rms_z} we show the calculated thermal noise as a function of redshift, at two different spatial scales.   Assuming the intrinsic signal has an RMS of order 10s of mK on these scales, we expect good signal to noise images up to $z\lesssim10$ and noise-dominated images from $z\gtrsim15$.

\subsubsection{Wedge removal}
Foregrounds represent one of the largest obstacles to detecting the 21 cm signal. 
Efforts generally focus on mitigating the foregrounds or discarding the Fourier modes expected to be dominated by foregrounds (see e.g. \citealt{kerrigan2018improved,chapman2019foregrounds}). 
Here we take the latter, conservative approach: excising a foreground-dominated \enquote{wedge} (in 2D cylindrical $k$-space) from the 21 cm lightcone.

An approximate relation for the contaminated wedge region \cite[e.g.][]{Morales2012,Vedantham2012,Trott2012,Parsons2014,Liu2014a,Liu2014b,wedge_murray} can be derived based on a baseline's response to a foreground point-source:
\begin{equation}
    k_{\|} \le \kappa(\b{k}_\perp, z) \equiv |\b{k}_\perp| \frac{E(z)}{1+z} \int_0^z \frac{dz'}{E(z')} \cdot \sin\theta + b \, ,
    \label{eq:horizon_cut_condition}
\end{equation}
where $E(z)=\sqrt{\Omega_m (1+z)^3 + \Omega_\Lambda}$, and $\b{k}_\perp$ are line-of-sight scales. 
Here, the additive term $b$ captures the fact that at low-$k_\perp$, the width of the foregrounds in Fourier space is constant, and set by an applied \enquote{frequency-taper}. We discuss this in more detail below. 
Finally, $\theta$ represents the zenith-angle of the point-source.  Since point-sources cover the sky, $\theta$ is defined by the viewing angle of the telescope.  Estimates of $\theta$ can range from the full-width-half-maximum of the telescope beam as an optimistic choice to the full horizon as a pessimistic choice \citep[cf.][]{Pober2014}.  We again make a conservative choice and assume the horizon limit, $\theta = \pi/2$.
Setting all Fourier modes that obey relation \ref{eq:horizon_cut_condition} to zero in principle removes all foregrounds (while at the same time removing some of the cosmic signal), and is known as \enquote{wedge removal}.

The wedge definition (Eq. \ref{eq:horizon_cut_condition}) is inherently redshift-dependent, and thus cannot be applied to the Fourier-transform of an entire lightcone, as the redshift evolves along the lightcone. To consistently define the wedge as a function of redshift, we implement a rolling procedure. For each redshift slice at comoving distance $r_{\|}$, we
\begin{enumerate}
    \item take the part of the lightcone in range $r_{\|} \pm \Delta r/2$, where $\Delta r = 750\, \mathrm{Mpc}$,
    \item multiply by Blackman-Harris (BH) taper function in the line-of-sight direction,
    \item Fourier-transform to 3D $\b{k}$-space and apply wedge removal,
    \item Transform back to real space and save values for the central slice only.\footnotemark
\end{enumerate}
Here, the BH taper is employed to reduce artificial ringing in Fourier space, which would otherwise occur when applying a simple box-car window function (i.e. just selecting a \enquote{chunk} of the lightcone) on a non-periodic boundary (e.g. \citealt{choudhuri2016visibility,trott2016chips}).
Such structure is inimical to the removal of foregrounds via the wedge, as the wedge relation (Eq. \ref{eq:horizon_cut_condition}) itself depends on a highly compact Fourier Transform of the foregrounds over the line-of-sight. High-$k_{||}$ ringing due to the window function leaks foreground power outside the wedge, reducing this leakage.

However, applying the BH taper is not without trade-offs. 
While it reduces Fourier-space sidelobes out to high-$k_{||}$, it increases the width of the main foreground lobe at low-$k_{||}$. This results in the buffer $b$, which does not evolve with perpendicular scale, but represents the minimum $k_{||}$ at any perpendicular scale for which the data is foreground-free. 
This parameter increases for more compact taper functions (in frequency space), and similarly for smaller \enquote{chunks} of the lightcone (i.e. $\propto \Delta r^{-1}$).
We define $b$ as the width in Fourier space where the dynamic range of the taper is $10^{-10}$. We found that $\Delta r = 750\,\mathrm{Mpc}$ optimizes between these trade-offs, but the results of this paper are not highly sensitive to the choice of it.

\footnotetext{If $M(\b{r})$ is the measured data, described rolling procedure amounts to
\begin{equation*}
    M^*(\b{r}_\perp, r_{\|}') = \mathcal{\widetilde{F}_{\b{r}}} \left\{
        W(\b{k}, r_{\|}') \cdot \mathcal{F_{\b{r}}} \left\{ 
            B_N(r_{\|} - r_{\|}') \cdot M (\b{r}) 
            \right\}
        \right\} \Big{|}_{r_{\|}'}\, ,
\end{equation*}
where $\mathcal{F_{\b{r}}}$ and $\mathcal{\widetilde{F}_{\b{r}}}$ are Fourier and inverse Fourier transforms, $W$ is the wedge window function computed from Eq. \ref{eq:horizon_cut_condition}, $B_N$ is the Blackman-Harris of size $N = \Delta r / 1.5 \, \mathrm{Mpc}$ centered around $r_{\|}'$.
}

\begin{figure*}
    \centering
    \begin{subfigure}[b]{\textwidth}
        \centering
        \includegraphics[width=\textwidth]{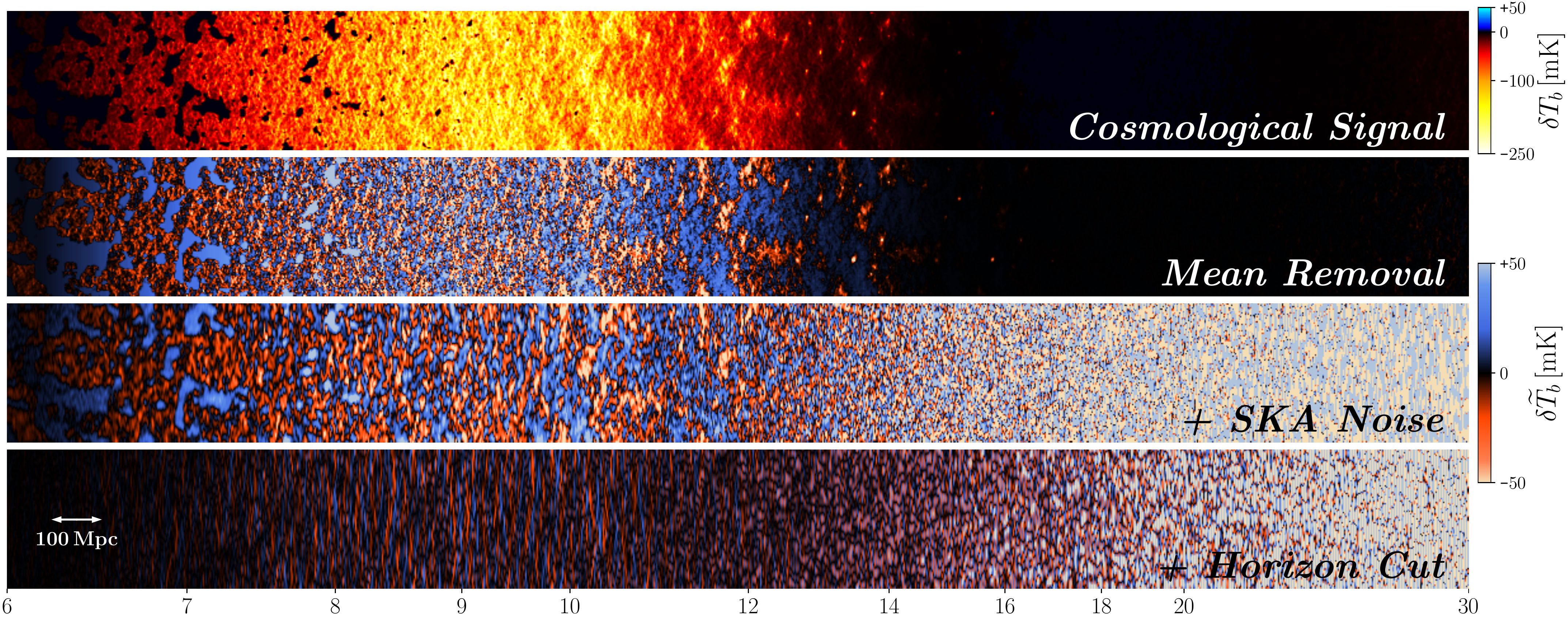}
        \caption{Slices through the frequency axis of an example 3D lightcone, following successive contamination of the signal. From top to bottom: \textit{Cosmological Signal}, \textit{Mean Removal}, \textit{Mean Removal + SKA noise}, \textit{Mean Removal + SKA noise + Horizon Cut}.}
        \vspace{0.2cm}
        \label{fig:lightcone_total_pipeline}
    \end{subfigure}
    \begin{subfigure}[b]{\textwidth}
        \centering
        \includegraphics[width=\textwidth]{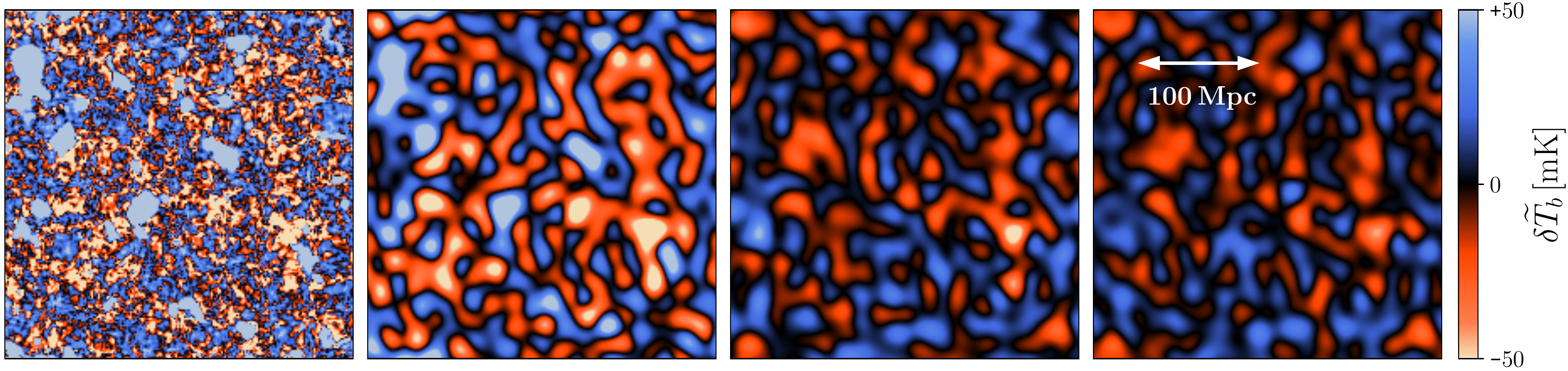}
        \caption{Slices trough the sky-plane at $z=8$, corresponding to the example in (a). From left to right: \textit{Mean Removal}, \textit{Mean Removal + SKA noise}, \textit{Mean Removal + SKA noise + Optimistic Wedge} ($\theta = 50^\circ$), \textit{Mean Removal + SKA noise + Horizon Cut} ($\theta = 90^\circ$).}
        \label{fig:lightcone_contaminated_zslice}
    \end{subfigure}
        \caption{Steps in our observational pipeline. We generate a database following each of the above three steps: \textit{Mean Removal}, \textit{+SKA Noise} and \textit{+Horizon Cut}.  In panel (a) we show slices along the frequency plane, while in panel (b) we show slices in the sky-plane.  In (b) we also include a more optimistic wedge contamination. Astrophysical parameters of the particular example are: $\zeta = 150$, $T_{\mathrm{vir}} = 10^{5.7} \, \mathrm{K}$, $L_X / \mathrm{SFR} = 10^{40} \, \mathrm{erg \, s^{-1} \, M_\odot^{-1} \, yr}$, $E_0 = 1.2 \, \mathrm{keV}$.}
        \label{fig:observation_pipeline}
\end{figure*}

\subsubsection{Dataset preparation}

Using the $10,000$ cosmological lightcone simulations described in Sect. \ref{sec:cosmo}, we generate three databases, one for each step in the observational pipeline outlined in Sect. \ref{sec:obs}.  From hereon, we refer to these three databases as
\begin{enumerate}
\item \textit{Mean Removal}
\item \textit{+SKA Noise}
\item \textit{+Horizon Cut}.
\end{enumerate}

Each database is randomly split as 80\%:10\%:10\% into training:validation:test sets.  In Figure \ref{fig:observation_pipeline}, we show an example 3D lightcone image (with cuts along the frequency (\ref{fig:lightcone_total_pipeline}) and sky-plane axes (\ref{fig:lightcone_contaminated_zslice})) resulting after each step of the signal contamination.  Removing the mean and adding SKA noise does not remove the large-scale features from the cosmological signal.  However, removing the modes from the wedge does contaminate the image in a way that is difficult to decipher by eye.  Below we quantify how well the NN recovers astrophysical parameters following each stage of contamination. Note that the wedge removal step could be considered pessimistic; current instruments like LOFAR and MWA are constantly improving foreground mitigation within the wedge \citep{li2018comparing,barry2019,LOFAR2020,hothi2021}. Moreover, NNs could be trained to add the missing, foreground-contaminated modes \citep{Gargon-Hartman21}. 
For a comparison, we calculate the wedge excision in a more optimistic case (see Figure \ref{fig:lightcone_contaminated_zslice}, where $\theta = 50^{\circ}$ is consistent with \citealt{LOFAR2020}), however we don't use it in the training of NNs.

For data augmentation, each lightcone is separated into $4$ smaller sky-plane patches and 10 realizations of noise are calculated per lightcone (see e.g. \citealt{perez2017effectiveness, shorten2019survey}). 
For computational efficiency, we downsample the 3D images by a factor of 4 (to voxels of 6 Mpc), noting that this is below the expected SKA1-low resolution.
Finally, to improve the stability in training, we normalize both the lightcone images and the parameter ranges to zero-mean and unit-variance (e.g. \citealt{lecun2012efficient}).

\section{Network Architectures} \label{ch:network_architectures}

Here we describe the different network architectures we use for parameter estimation.  These are based on combinations of convolutional and recurrent operations, guided by our specific usage case.  Namely, the cosmic 21 cm signal contains
spatially-correlated information in the sky-plane and spatio-temporally correlated information across frequency bins. These correlations, intrinsic to the cosmic signal, are however weakened with increased data contamination from observational pipelines. 

With this in mind, we construct three NN architectures - all using (primarily) $2\mathrm{D}$ convolutions to encode local sky-plane correlations and either recurrent or $1\mathrm{D}$ convolutional operations to encode information along the frequency dimension.
Although CNNs can 
encode correlations in both sky and frequency planes, they treat the data as a stationary image. On the other hand, RNNs are designed to encode a sequence of data by \enquote{rolling over} it and reusing the same weights on each step (see Appendix \ref{app:layers} for more details). One RNN layer then effectively becomes a NN with the depth equal to the length of a sequence - able to encode highly non-linear data by having comparably fewer weights. This design allows RNNs, and especially Long Short Term Memory RNNs \citep{LSTM}, to efficiently and quickly find a stable local minimum of the loss function when training.\footnote{Global minima of loss functions are almost impossible to find given the high dimensionality of the parameter space of deep network weights; however, local minima can result in satisfactory, comparable performance (e.g. \citealt{choromanska2015}).} 
As a result, RNNs became famous in audio/video encoding (e.g. \citealt{ConvLSTM,zhao2019hierarchical}) and natural language processing (e.g. \citealt{aharoni2018gradual}). Here we introduce them to the field of 21 cm.

Below, we briefly sketch the specific CNN and RNN architectures we use in this study.  Detailed descriptions, including the number of layers, filters, etc. can be found in Appendix \ref{app:DetailedArchitectures}.

\subsection{CNN}
\begin{figure}
    \centering
    \includegraphics[width=\linewidth]{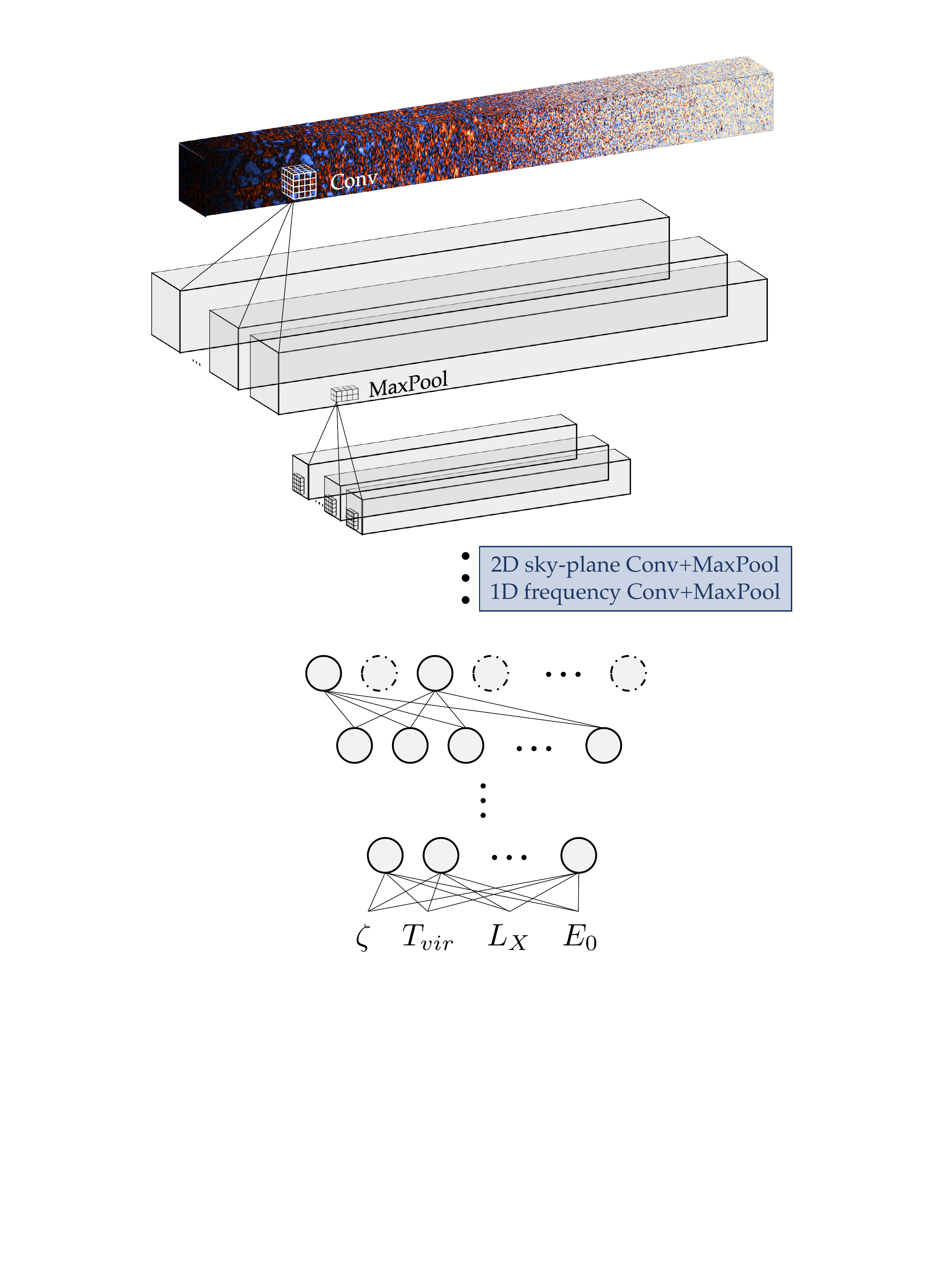}
    \caption{CNN architecture sketch. Parallel lightcones denote multiple filters and subsequent convolutional and pooling layers are depicted in blue. More details are found in Appendix \ref{app:DetailedArchitectures}.}
    \label{fig:CNN}
\end{figure}

Figure \ref{fig:CNN} shows a sketch of our CNN architecture.  The convolutional part consists of iterative convolutional (\enquote{Conv}) and pooling (\enquote{MaxPool}) layers, and is followed by fully connected (FC) layers.  The final output is the prediction of the four astrophysical parameters we use in this study.  \enquote{Neurons} are shown with circles, where dash-dotted circles depict dropout at the first FC layer. Convolutions locally correlate voxels of the lightcone and pooling layers downsample it by keeping only the strongest activations.

In the first convolutional + pooling layer we use 3D kernels. Subsequently, we iterate successive layers of $2\mathrm{D}$ convolutions in the sky-plane and $1\mathrm{D}$ convolutions across frequency bins. Two such layers combined effectively make a $3\mathrm{D}$ convolutional layer, however with a reduced number of weights.

One advantage of CNN architectures is that they are comparably simple, capable of being trained using more modest computational resources.  Our CNNs on average require 0.13 GPUh (NVIDIA P100)
per epoch of training, and 20 ms per execution once trained.

\subsection{ConvRNN}

\begin{figure}
    \centering
    \includegraphics[height=1.5\linewidth]{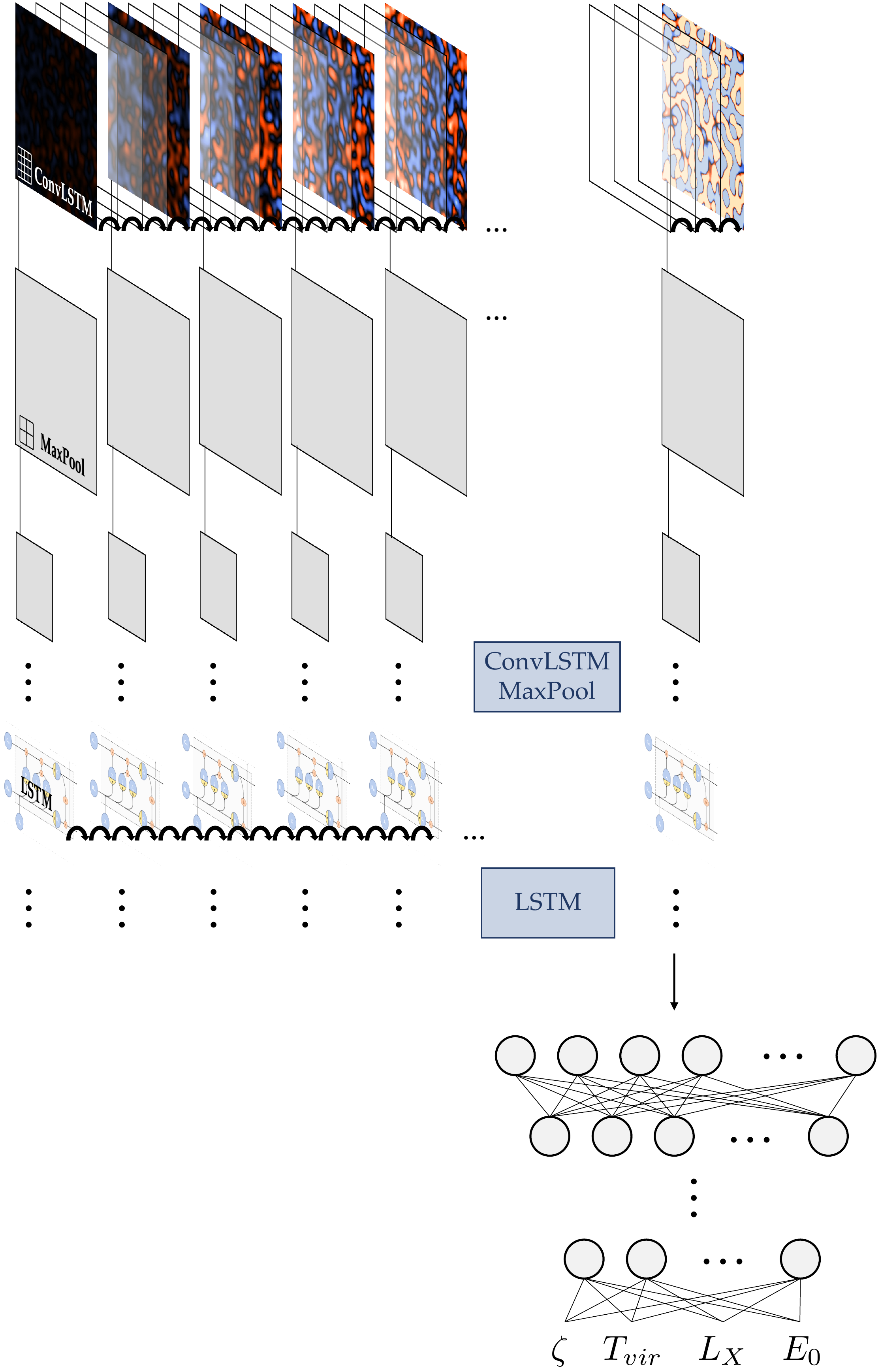}
    \caption{ConvRNN architecture sketch. For legibility, multiple filters/channels are omitted from the sketch. Curly arrows represent activations passing sequentially in the LSTM layer. Consecutive layers are depicted in blue. More details are found in Appendix \ref{app:DetailedArchitectures}.}
    \label{fig:ConvLSTM}
\end{figure}

Figure \ref{fig:ConvLSTM} shows a sketch of our ConvRNN architecture. The input lightcones are passed through a series of convolutional Long Short Term Memory (ConvLSTM; \citealt{ConvLSTM}) layers, which combine $2\mathrm{D}$ convolutions in the sky-plane with recurrent LSTMs in the frequency dimension. As for the CNN, the dimensionality is reduced following each convolution with a MaxPool layer.  The ConvLSTM + MaxPool layers are then followed by pure LSTM and finally FC layers, leading to the parameter predictions. 

Despite their efficient performance, a notable drawback is the requirement of substantial computational resources.  Our ConvRNNs on average require 2 GPUh (NVIDIA P100) per epoch of training, and 6 s per execution once trained.  This is a factor of 15 larger in the training time compared with the CNN described above.  Because of these substantial computational resources, we also implement a ``slimmed-down'' RNN that we refer to as ``SummaryRNN'' below.

\subsubsection{SummaryRNN}

The main computational bottleneck of the ConvRNN is the convolutional back propagation through time (e.g. \citealt{BPTT}). Therefore we also construct a ``SummaryRNN'', in which we remove all recurrent operations from the convolutions, replacing the ConvLSTM layers in Fig. \ref{fig:ConvLSTM} with pure $2\mathrm{D}$ sky-plane convolutions. Thus, the SummaryRNN first ``summarizes'' the lightcones using sequential sky-plane convolutions, and only passes these (downsampled) summaries into a stack of LSTM and FC layers.  The SummaryRNN architecture sketch would effectively be identical to the ConvRNN in Fig. \ref{fig:ConvLSTM}, but without the curved arrows (representing recurrent operation) in the upper row.

Although the resulting SummaryRNN loses some information compared with the ConvRNN, it can train considerably faster.  Specifically, our SummaryRNNs take 0.1 GPU/h (NVIDIA P100) per epoch, and 45 ms to execute.  This is a factor of 20 improvement in training time compared with ConvRNN.

\section{Training} \label{sec:training}

\subsection{General network set-up} \label{subsec:network_setup}

We make use of several standard techniques to improve the stability and generalization of the NN training, including optimal initialization of the network, batch normalization and dropout. 
For each choice of activation function, we use the corresponding optimal initialization schemes identified in \citet{lecun2015deeplearning} and \citet{he15} on the basis of keeping the variance of the weights constant during the training. The activation function is applied to all hidden layers of the network. However for LSTMs, we keep the internal structure and activations unchanged (for details see Fig. \ref{fig:LSTM_cell}).  Dropout \citep{srivastava14} is applied only once, in the widest part of every NN. The performance of the network is evaluated by calculating the mean square error (MSE) between the true and predicted values of the parameters. The weights are updated with back-propagation using a fixed batch size \citep{10.5555/65669.104451}. Batch normalization is applied immediately before any tensor transformation, with an exception of a few layers at the very end \citep{ioffe15}.

We perform a grid search of standard NN hyperparameter combinations, listed in Table \ref{table:hyperaparams}.  For this, we train each NN over a reduced number of epochs (100) and use the resulting MSE loss to identify the best hyperparameters.  This required $\approx 500$ trainings per architecture and database.  
Hyperparameters marked with red in Table \ref{table:hyperaparams} are preferred by all combinations of NN architectures and databases.
In general, the performance was extremely sensitive to the initial learning rate. We notice better performance for lower batch size and dropout, while still allowing the final network to generalize well (as we shall see in the next section).  Turning on batch normalization made the training more stable, while we saw no difference between \texttt{Adam} and \texttt{Nadam} optimizers, and between \texttt{relu} and \texttt{leakyrelu}\footnote{
Here we use 
$
    \texttt{leakyrelu(0.1)} = \begin{cases}
    0.1 \, x & x < 0\\
    x & x \ge 0
    \end{cases} \, .
$}
activations. Final architectures are then trained until convergence (for a detailed specification, we refer the reader to Appendix \ref{app:DetailedArchitectures}).

The training was done in parallel on $10$ NVIDIA P100 GPUs with the \texttt{ring-Allreduce} update scheme. For this, we used \texttt{TensorFlow}\footnote{\url{https://github.com/tensorflow/tensorflow}} \citep{tensorflow2015-whitepaper} as a main framework and \texttt{horovod}\footnote{\url{https://github.com/horovod/horovod}} \citep{horovod} as a parallelization library. 
When training on multiple GPUs, a copy of a NN is held on each device and the final loss is an average across all individual losses. Thus, with the constant batch size ($\mathrm{bs}$) per device, the effective batch size grows with the number of GPUs ($N_\mathrm{GPU}$) as $\mathrm{bs_{eff}} = N_{\mathrm{GPU}} \cdot \mathrm{bs}$. To cope with this, in the first $10$ epochs we linearly scale the learning-rate $\mathrm{lr} \rightarrow \mathrm{lr} \cdot N_{\mathrm{GPU}}$ (so-called \enquote{warmup}, see \citet{warmup} for details).
 
Finally, for a consistent comparison, we fixed the learning rate scheduler -- reducing it by a factor of $10$ on $50\%$ and $75\%$ of the training.

\begin{table}
\centering
\begin{tabular}{|r|c|} 
 \hline
 Batch Size & \textcolor{red}{$20$}, $100$ \\
 \hline
 Initial Learning Rate & $10^{-2}$, \textcolor{red}{$10^{-3}$}, $10^{-4}$, $10^{-5}$ \\
 \hline
 Dropout & \textcolor{red}{$0.2$}, $0,5$ \\
 \hline
 Batch Normalization & \textcolor{red}{\texttt{True}}, \texttt{False} \\
 \hline
 Optimizer & \texttt{RMSprop}, \texttt{SGD}, \texttt{Adamax}, \textcolor{red}{\texttt{Adam}}, \textcolor{red}{\texttt{Nadam}} \\
 \hline
 Activation Function & \textcolor{red}{\texttt{relu}}, \textcolor{red}{\texttt{leakyrelu(0.1)}}, \texttt{elu}, \texttt{selu} \\
 \hline
\end{tabular}
\caption{Hyperparameter space explored with a grid search.  Training was performed using a reduced number of epochs for various hyperparameter combinations. The final choices are indicated in red.  For more details on the network architectures, see Appendix \ref{app:DetailedArchitectures}.}
\label{table:hyperaparams}
\end{table}

\subsection{Training performance} \label{subsec:training_performance}

\begin{figure*}
    \centering
    \includegraphics[width=\linewidth]{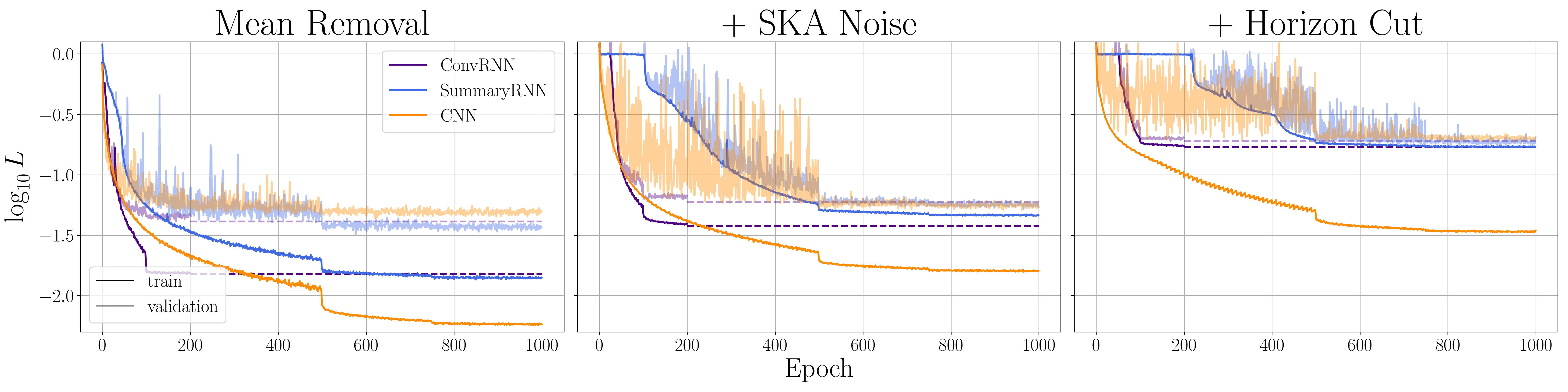}
    \caption{Training and validation losses for the final NN architectures.  CNN, ConvRNN and SummaryRNN architectures are depicted in orange, purple and blue, respectively.  The panels correspond to our three databases, from left to right in increasing levels of signal contamination. ConvRNN converges extremely quickly and we train it for 200 epochs only (dashed lines denote the final loss after 200 epochs). The other NNs were trained for 1000 epochs.}
    \label{fig:all_losses}
\end{figure*}

In Figure \ref{fig:all_losses} we show the training and validation losses for our three network architectures.  The panels correspond to our three databases: (i)~{\it Mean removal}; (ii)~{\it +~SKA noise}; and (iii)~{\it +~Horizon cut}, from left to right.  For the {\it Mean removal} database, we see that both RNNs outperform the CNN.  This is evidenced mainly by their validation losses being lower.  Furthermore, the difference between the training and validation losses is much smaller for the two RNNs than for the CNN, demonstrating that the RNNs are able to generalize better and are less prone to over-fitting. 

We see that the final validation losses of ConvRNN and SummaryRNN are comparable for the {\it Mean removal} database.  However, the ConvRNN (which includes recurrent layers also in the convolutional steps) is much more stable in training.  It rapidly and with very little stochasticity finds a local minimum in the loss function,
after only 100 epochs; SummaryRNN requires 500 epochs to approach a comparable loss and the training is noisier initially (note the effect of LR reduction at 50\% and 75\% of the training).  However, even accounting for $\sim5$ times more training epochs, SummaryRNN still is less computationally intensive compared to ConvRNN (a factor of 4 fewer GPUh in total training time).

From the middle and right panels of Fig. \ref{fig:all_losses} we see that with higher contamination, the performance worsens and the differences in the final validation losses between the architectures disappear.  The fact that the different architectures are reaching the same validation loss is strongly suggestive that we are reaching the intrinsic limits of our datasets.

\begin{figure}
    \centering
        \includegraphics[width=\linewidth]{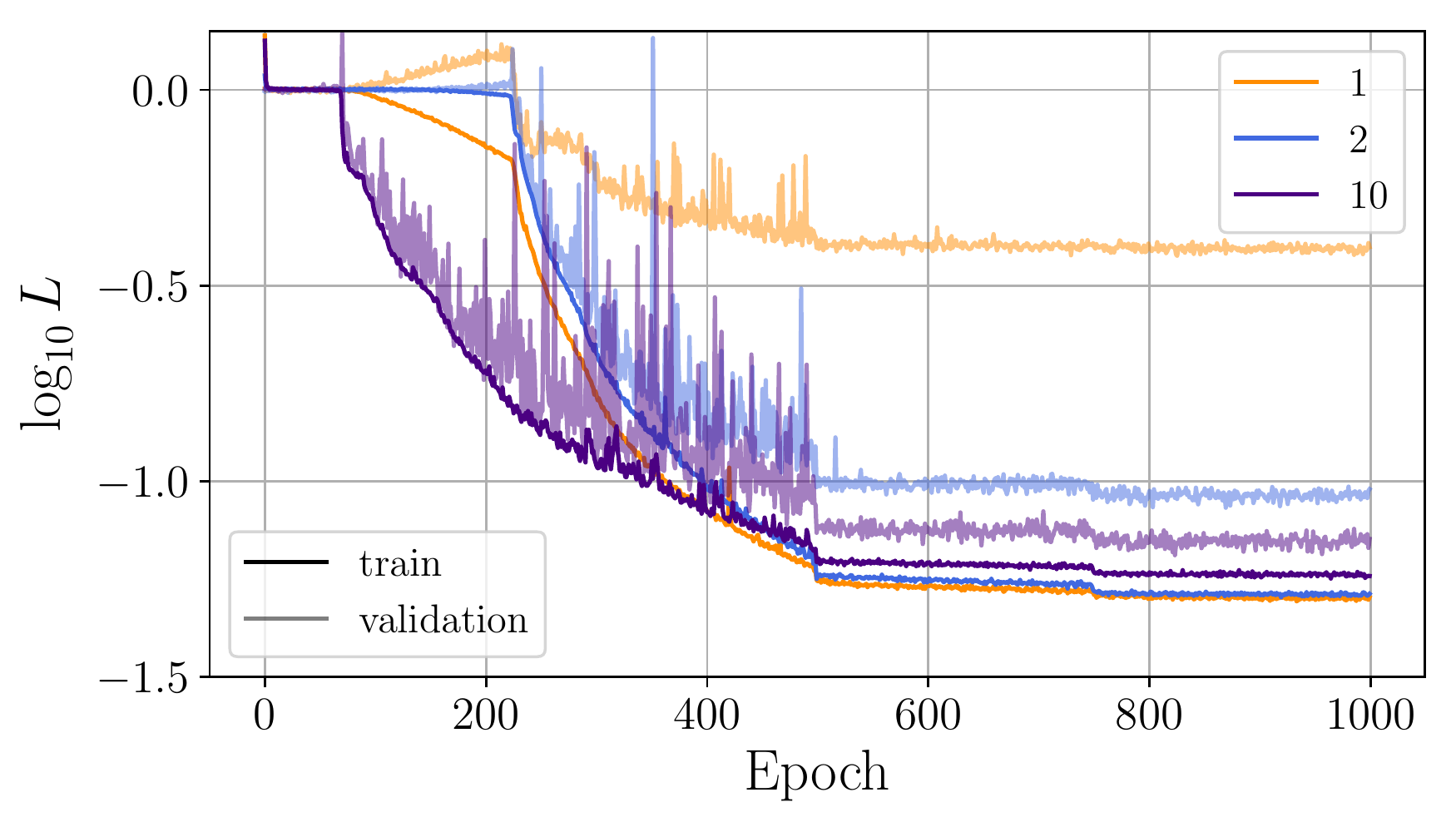}
        \caption{Evolution of the training and validation losses for SummaryRNN when varying the number of noise realizations per cosmic signal in the {\it + SKA noise} database.  The fiducial databases used here include 10 realizations per cosmic signal.  We see that the validation loss keeps decreasing as more noise realizations are included.  This is suggestive that our final NN performance is limited by our data augmentation (database size).}
        \label{fig:noise_augmentation}
\end{figure}

To explore this point further, we vary the number of noise realizations per cosmic signal in the {\it + SKA noise} database, re-training the SummaryRNN each time.  In Figure \ref{fig:noise_augmentation} we show the training and validation losses using 1, 2 and 10 noise realizations per cosmic signal (our fiducial database corresponds to 10).  We note a significant decrease in the validation loss going from 1 to 2 noise realizations per cosmic signal.  However, although the improvement is smaller, the final validation loss keeps decreasing even down to 10 realizations.  This supports the claim above that our results are limited by our data augmentation, especially for the {\it + SKA Noise} and {\it + Horizon cut} databases.  In future work, we will increase the size of the databases, sampling more cosmic signals and contamination realizations, quantifying if we reach convergence.

\section{Parameter recovery} \label{sec:parameter_recovery}

\begin{figure*}
    \centering
    \includegraphics[width=0.99\linewidth]{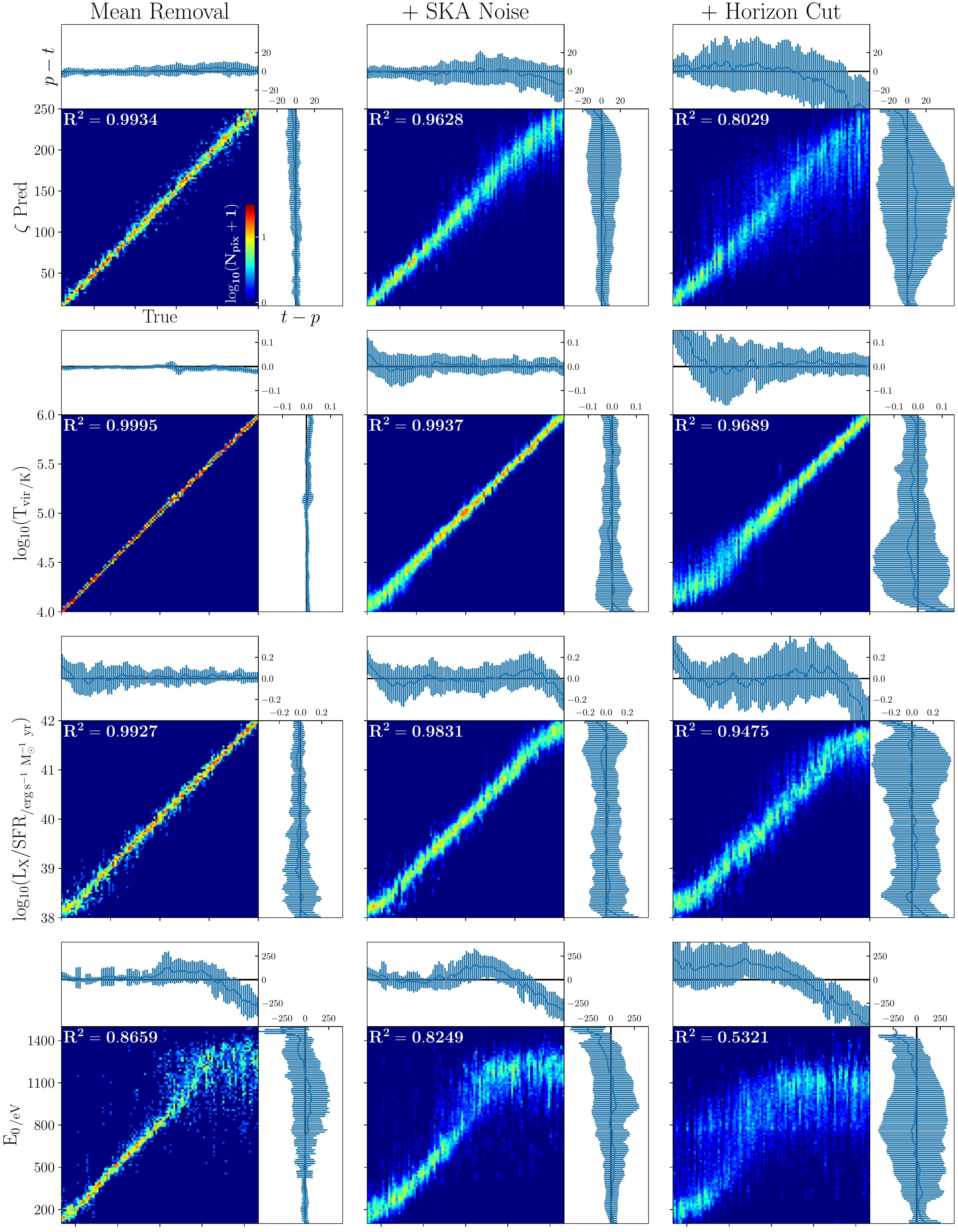}
    \caption{Histograms of True vs. Predicted, $P(p,t)$, for our SummaryRNN architecture and all databases. From left to right, columns represent: \textit{Mean Removal}, \textit{+SKA noise}, \textit{+Horizon Cut} datasets. Rows correspond to our four astrophysical parameters (defined in the text). Each panel is bordered by the mean and $\pm 1 \sigma$ of the 1D conditional distributions, $P(t-p|p)$ and $P(p-t|t)$. $R^2$ coefficients are marked in upper-left corners. Some labels are omitted to avoid overcrowding; see the top-left panel for the relevant labels.}
    \label{fig:true_vs_pred}
\end{figure*}

In this section we demonstrate the performance of the trained networks in parameter recovery on the test sets.  We begin by showing the predicted vs. true distributions, $P(p,t)$, for the SummaryRNN in Figure \ref{fig:true_vs_pred}.  The rows correspond to our four astrophysical parameters, while the columns correspond to the three different databases.  Standard $R^2$ scores are reported in the panels. On the top and side of each 2D distribution we show the mean and $\pm1\sigma$ of the 1D conditional distributions, $P(t - p | p)$ and $P(p - t | t)$, respectively.

Focusing on the {\it Mean Removal} results, we recover the same qualitative trends discussed in \citet{gillet19}, who used the same database of cosmic signals.  The best recovery is obtained for the minimum virial temperature for star forming galaxies, $T_{\rm vir}$, with an $R^2 = 0.9995$ and no notable biases. This is because $T_{\rm vir}$ impacts the timing of {\it all} astrophysical epochs, as well as the characteristic scales of structures.  Thus the cosmic signal is very sensitive to $T_{\rm vir}$, facilitating good recovery.  The ionizing efficiency and X-ray luminosity parameters, $\zeta$ and $L_{X}/{\rm SFR}$, are also predicted very well.  These impact the timing of the EoR and the EoH, respectively.  The minimum X-ray energy escaping the host galaxy, $E_0$, is recovered well for values below $E_0 < 1$ keV.  Since the interaction cross section for X-rays is a strong function of energy, photons with higher energies are inefficient at ionizing / heating the IGM and do not leave a strong imprint in the 21 cm signal.  As we approach $E_0 \rightarrow 1.5$ keV, corresponding to photons with mean free paths comparable to the Hubble length,  the network prediction becomes understandably randomized (pulling towards the mean of the range).  Indeed, all distributions show a characteristic \enquote{S} shape, as a consequence of the sharp boundary on the parameter ranges.

As could be expected from the validation loss curves in the previous section, the SummaryRNN predictions on the test sets notably worsen with increasing signal contamination (going from left to right in Fig. \ref{fig:true_vs_pred}).  Including SKA noise\footnote{We remind the reader that our noise calculation is done in $uv$ space, and includes the effects of the finite beam.} only decreases the $R^2$ scores by a few percent.  The wedge excision has a more dramatic effect, especially on the $\zeta$ and $E_0$ parameter predictions which drop to $R^2=$ 0.80 and 0.53, respectively.  

\begin{figure*}
    \centering
    \includegraphics[width=\linewidth]{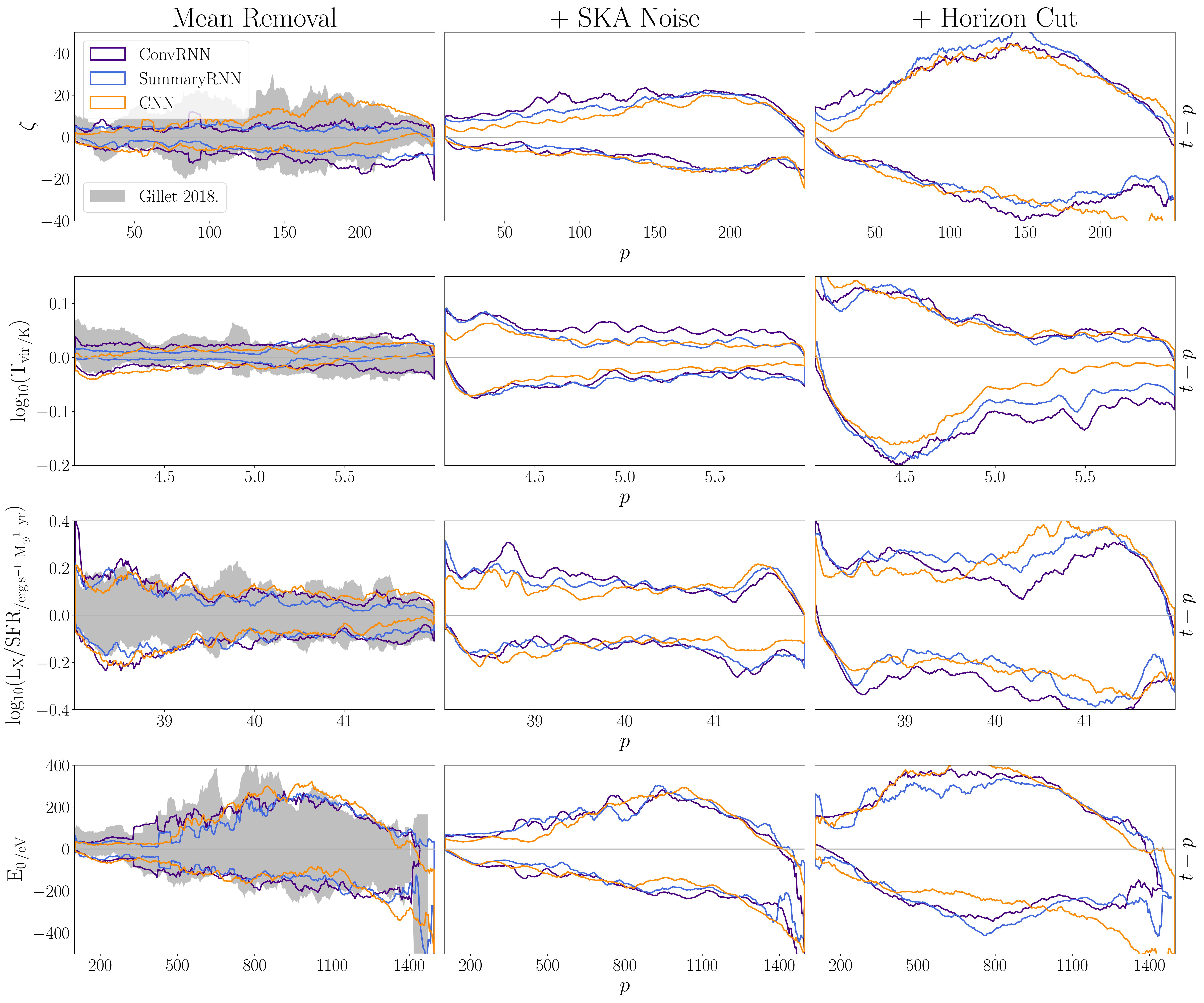}
    \caption{Prediction errors for all architectures.  The lower and upper set of corresponding curves bracket $\pm1\sigma$ of the $P(t-p|p)$ distributions.  Columns represent different test databases, and rows different astrophysical parameters.}
    \label{fig:all_best}
\end{figure*}

After showing all of the predicted vs. true distributions for the SummaryRNN architecture, we now compare the parameter recovery from all three of our architectures, using their $P(t|p)$.\footnote{Although $P(p|t)$ is a common performance metric in the literature, it is less meaningful than $P(t|p)$.  In practice we will have an observation that when fed into a trained network will give a \enquote{best guess} parameter vector, $\boldsymbol{\theta^p}$.  So the relevant uncertainty is the probability of the \enquote{true} parameters of the Universe, given this network prediction, $P(\boldsymbol{\theta^t} | \boldsymbol{\theta^p})$.  Unfortunately, our point estimate networks do not allow for a direct calculation of the Bayesian posterior; however, the closest approximation we can make with the data at hand are the \enquote{marginalized} distributions of $P(t|p)$ from the test set, where $t$ and $p$ are components of $\boldsymbol{\theta^t}$ and $\boldsymbol{\theta^p}$ respectively.
}
In Figure \ref{fig:all_best}, we bracket $\pm 1 \sigma$ (RMS) of the $P(t-p|p)$ 
 distributions for all architectures (identical to $P(t|p)$, only shifted for easier visualization), with upper and lower curves of matching colors. Columns represent different databases and rows different astrophysical parameters. In the case of unbiased errors, $P(t-p|p)$ would be a zero-mean distribution.

As expected from the validation loss curves, the two RNNs perform the best on the {\it Mean Removal} test data.  Overall, SummaryRNN performs the best, likely outperforming ConvRNN due to the longer training (the number of training epochs for ConvRNN was five times less than for SummaryRNN, due to its substantial computational requirements, as discussed above).  Specifically, SummaryRNN has MSE on average a factor of $\sim2$ lower than the CNN for all parameters. Here the MSE is calculated for each parameter individually, averaging over all test samples. For detailed numerical values for all architectures and parameters, see Table \ref{table:variances}.
\begin{table}
\centering
\begin{tabular}{c||c|c|c|} 
  &  Mean Rem. & +SKA Noise & +Hor. Cut \\
 \hline\hline
 $\zeta$ 
 & \begin{tabular}{@{}c@{}} \cCRNN{$65$} \\ \cSRNN{$31$} \\ \cCNN{$71$} \end{tabular}
 & \begin{tabular}{@{}c@{}} \cCRNN{$207$} \\ \cSRNN{$175$} \\ \cCNN{$144$} \end{tabular}
 & \begin{tabular}{@{}c@{}} \cCRNN{$921$} \\ \cSRNN{$925$} \\ \cCNN{$879$} \end{tabular}\\
 \hline
 $log_{10}(T_\mathrm{vir})$
 & \begin{tabular}{@{}c@{}} \cCRNN{$58 \cdot 10^{-5}$} \\ \cSRNN{$16 \cdot 10^{-5}$} \\ \cCNN{$35 \cdot 10^{-5}$} \end{tabular}
 & \begin{tabular}{@{}c@{}} \cCRNN{$260 \cdot 10^{-5}$} \\ \cSRNN{$198 \cdot 10^{-5}$} \\ \cCNN{$125 \cdot 10^{-5}$} \end{tabular}
 & \begin{tabular}{@{}c@{}} \cCRNN{$1175 \cdot 10^{-5}$} \\ \cSRNN{$981 \cdot 10^{-5}$} \\ \cCNN{$749 \cdot 10^{-5}$} \end{tabular}\\
 \hline
 $log_{10}(L_X/SFR)$
 & \begin{tabular}{@{}c@{}} \cCRNN{$15 \cdot 10^{-3}$} \\ \cSRNN{$9 \cdot 10^{-3}$} \\ \cCNN{$12 \cdot 10^{-3}$} \end{tabular}
 & \begin{tabular}{@{}c@{}} \cCRNN{$24 \cdot 10^{-3}$} \\ \cSRNN{$22 \cdot 10^{-3}$} \\ \cCNN{$17 \cdot 10^{-3}$} \end{tabular}
 & \begin{tabular}{@{}c@{}} \cCRNN{$77 \cdot 10^{-3}$} \\ \cSRNN{$68 \cdot 10^{-3}$} \\ \cCNN{$64 \cdot 10^{-3}$} \end{tabular}\\
 \hline
 $E_{0}$
 & \begin{tabular}{@{}c@{}} \cCRNN{$25 \cdot 10^{3}$} \\ \cSRNN{$22 \cdot 10^{3}$} \\ \cCNN{$28 \cdot 10^{3}$} \end{tabular}
 & \begin{tabular}{@{}c@{}} \cCRNN{$30 \cdot 10^{3}$} \\ \cSRNN{$28 \cdot 10^{3}$} \\ \cCNN{$30 \cdot 10^{3}$} \end{tabular}
 & \begin{tabular}{@{}c@{}} \cCRNN{$77 \cdot 10^{3}$} \\ \cSRNN{$82 \cdot 10^{3}$} \\ \cCNN{$81 \cdot 10^{3}$} \end{tabular} \\
 \hline
\end{tabular}
\caption{MSE of all considered models and parameters. All values are expressed in parameter units squared (e.g. $E_{0}$ in $\mathrm{eV}^2$). ConvRNN, SummaryRNN and CNN are marked with purple, blue and orange, respectively.}
\label{table:variances}
\end{table}

In these {\it Mean Removal} panels of Fig. \ref{fig:all_best}, we also show in gray the corresponding limits from the CNN presented in \citet{gillet19}.  Training on the same database of cosmic signals, \citet{gillet19} considered a shallower CNN than we use here, and did not include any signal contamination (not even mean removal).  Our deeper networks result in a factor of $\sim 2-8$ smaller variance in $P(t|p)$ (calculated as an average across all test samples), despite having a factor of 4 times poorer resolution and having removed the mean of the cosmic signal.

For the \textit{+ SKA Noise} test database, the prediction errors are larger for all parameters, and there are little differences between network performances (suggestive that we are limited by the dataset size, as discussed previously).  Some of the qualitative trends are understandable on physical grounds.  As noted previously, for high values of $E_0$ that do not impact the signal the network predictions are almost randomly distributed across the whole range, resulting in the negative bias seen in the figure beyond $E_0\gsim$ 1.2 keV. Furthermore, for low values of $T_{\mathrm{vir}}$, the first galaxies form at very high redshifts, shifting all astrophysical epochs of the 21 cm signal to frequencies with higher thermal noise.  This explains the (modest) increase in the prediction error at $T_{\rm vir} \lsim 10^{4.3}$ K.  The fact that the increase is relatively modest suggests all networks have learned to \enquote{marginalize over} the noise reasonably well even with only 10 noise realizations per cosmic signal.

For the \textit{+ Horizon Cut} test database, the prediction errors increase significantly for all architectures.  This is not surprising given that our pessimistic wedge removal throws away a significant amount of information (c.f. Fig. \ref{fig:observation_pipeline}).  Unlike for the thermal noise that is uncorrelated with the cosmic signal, we cannot augment our \textit{+ Horizon Cut} database without running more samples of the cosmic signal.  As a result, all networks are unable to generalize and perform much worse than with the other databases.  In future work, we will investigate how well NNs can train to marginalize over simulated foregrounds, rather than adopting the simple foreground-avoidance approach as we do here (for a foreground-cleaning example, see \citealt{21cm_ML_LaPlante19}).

\subsection{What features are guiding the network predictions?} \label{subsec:visualization}

One major drawback of machine learning is that deep neural networks are often treated as \enquote{black boxes}.  As such, it is important to check if the trained NN is using reasonable features in the images to make predictions, and is not overfitting by focusing on unphysical artifacts particular to a dataset (e.g. \citealt{lapuschkin2019unmasking}).  For this reason, feature identification tools such as saliency mapping and attention mechanisms are becoming increasingly popular (e.g. \citealt{zeiler2014visualizing,selvaraju2017grad,chang2018explaining,vaswani2017attention,ramachandran2019stand}).

\begin{figure*}
    \centering
    \includegraphics[width=\linewidth]{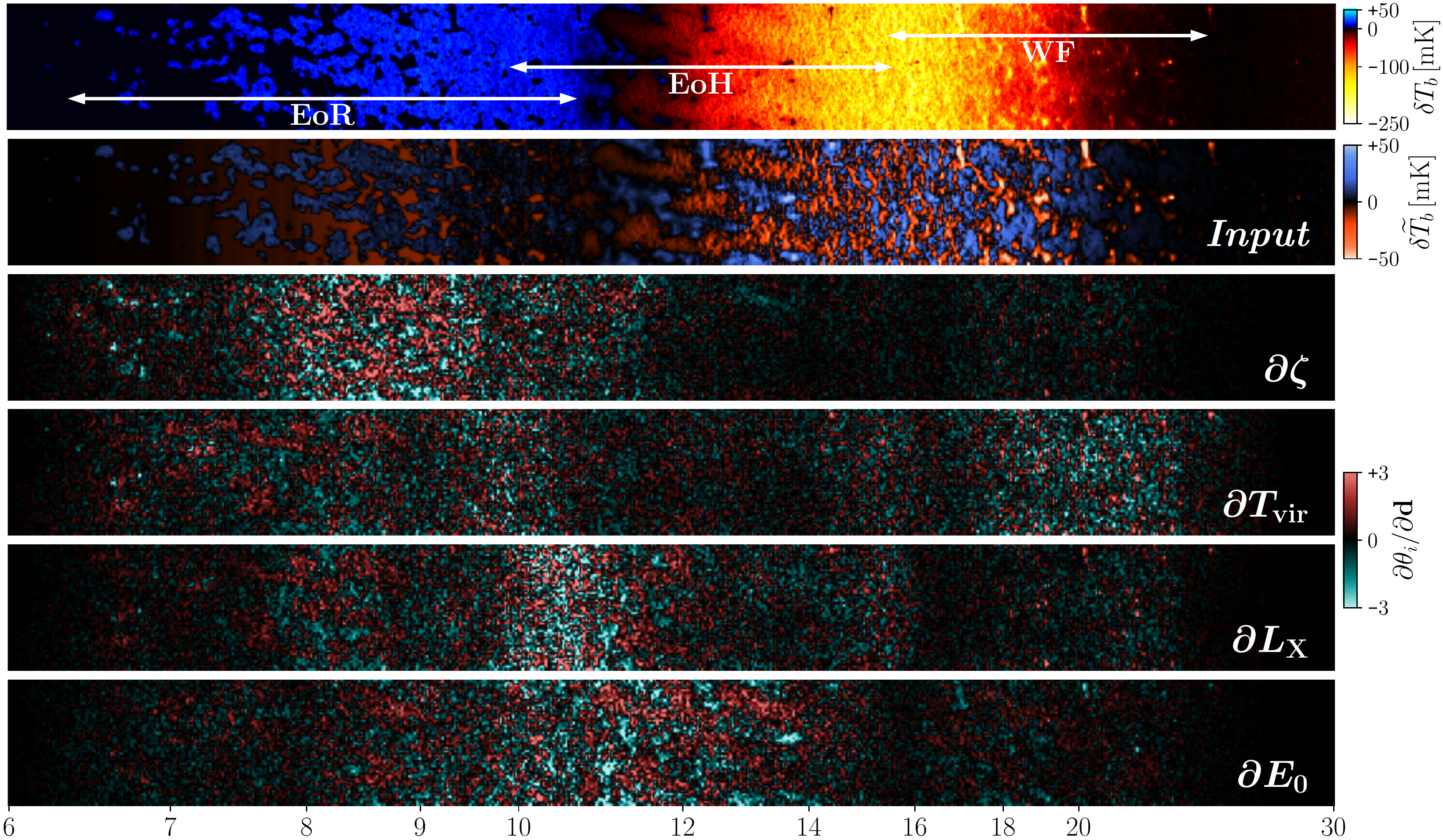}
    \caption{Gradient saliency maps for a sample image from the \textit{Mean Removal}  test set, constructed with our trained CNN. The top panel corresponds to the cosmic signal, the second panel to the input image, and the bottom four are the gradient saliency maps for the indicated parameter (normalized to have unit variance). For visualization purposes, we only show 2D slices through the 3D lightcones. The brighter colors (larger gradients) highlight the regions of the image that are important for each parameter prediction.  For each  parameter, the network correctly focuses on the relevant astrophysical epoch(s) of the 21 cm signal (denoted with arrows in the top panel).
    For reference, the cosmic signal used in this example corresponds to the following parameters: $\zeta = 44$, $T_{\mathrm{vir}} = 10^{4.7} \, \mathrm{K}$, $L_X / \mathrm{SFR} = 10^{40} \, \mathrm{erg \, s^{-1} \, M_\odot^{-1} \, yr}$, $E_0 = 0.6 \, \mathrm{keV}$.}    \label{fig:jacobian_CNN}
\end{figure*}

Here we use a simple saliency mapping technique \citep{simonyan13} to visualise the features used by our CNN.\footnote{Unfortunately, RNN visualization requires more complicated techniques \citep{karpathy15, ramanishka17, bargal18} that are not straightforward to adapt to our usage case.  We thus use the CNN feature maps as a \enquote{sanity check} in this section.} 
Specifically, we calculate a gradient saliency map, constructing a Jacobian matrix of the NN prediction with respect to the input image:
\begin{equation}
\label{eq:grad}
    \mathcal{J}_0 = \left\langle\frac{\partial \boldsymbol{\theta^p}}{\partial \b{d}} \Big{|}_{\b{d}_0}\right\rangle \, ,
\end{equation}
where $\boldsymbol{\theta^p}$ is the NN parameter prediction vector in which each component corresponds to an astrophysical parameter, $\b{d}$ is the data vector in which each component corresponds to a given pixel of the 3D lightcone image, $\b{d}_0$ is a given input image for which we evaluate the saliency map, and the averaging is performed over all possible $150\times150 \,\mathrm{Mpc}$ sky-plane cuts out of the initial $300\times300 \,\mathrm{Mpc}$ lightcone.
Besides giving us a consistent way to visualize a saliency map for the whole simulation volume, averaging smooths-out fluctuations in the gradients (for details about gradient smoothing, see \citealt{smilkov17}).
Intuitively, the gradient $\partial \boldsymbol{\theta^p}/\partial \b{d}$ corresponds to the change in the NN parameter prediction from a pixel-by-pixel perturbation in the input image. 

In Figure \ref{fig:jacobian_CNN} we use a sample input from our \textit{Mean Removal} test set to illustrate the feature identification of our CNN.  For simplicity, we only show 2D slices through the 3D lightcones.  The top panel corresponds to the cosmic signal, the second panel to the input image, and the bottom four are the (normalized) gradient saliency maps for each parameter.  These gradient saliency maps are also commonly referred to as \enquote{heat maps}, since brighter colors (larger gradients) highlight the regions of the image that are important for each parameter prediction.

Unlike everyday images, our lightcones do not have \enquote{objects} (cats, dogs, boats...) that would make feature identification from heat maps straightforward.  However, we can still draw some general conclusions from Fig. \ref{fig:jacobian_CNN}, by noting that the network has learned to correctly focus on the astrophysical epoch(s) relevant to each parameter (see the discussion of parameters in Section \ref{sec:cosmo}).  For example, the minimum virial temperature for star-forming galaxies, $T_{\rm vir}$, impacts when galaxies form.  Thus it drives the timing of {\it all} astrophysical epochs: EoR, EoH, Wouthuysen–Field (WF) coupling (roughly demarcated with arrows for this example in the top panel of Fig. \ref{fig:jacobian_CNN}).  Indeed, the gradient saliency map shows that the $T_{\rm vir}$ prediction is sensitive to the whole redshift range of the input image.  Conversely, the ionizing efficiency, $\zeta$, is mostly relevant for the timing of the EoR: the NN correctly focuses on this epoch, additionally using the \textsc{Hi} patches remaining during the late stages ($z\sim7$ for this model) for its prediction.  Likewise, the X-ray heating parameters, $L_X/$SFR and $E_0$, are most sensitive to the EoH.  While $L_X/$SFR mostly regulates the relative timing of the EoH, $E_0$ parametrizes the hardness of the emerging X-ray spectra and thus drives the relative sizes of the heated regions.  Indeed, we see in the heat maps that $E_0$ is more sensitive to changes in the large, heated IGM structures.

We conclude therefore that the qualitative trends shown in Fig. \ref{fig:jacobian_CNN} provide a good sanity check that our NN has learned physically-relevant information and is not overfitting. Drawing more quantitative insights from gradient saliency maps is difficult, given that they only show parameter sensitivity to uncorrelated, pixel-by-pixel perturbations.  Saliency maps using different basis sets (e.g. wavelets, superpixels), might be more useful for physically-meaningful feature identification from 21 cm lightcones. We defer this to future work.

\section{Conclusions} \label{ch:discussion_and_conclusions}

Upcoming images of the cosmic 21 cm signal will provide a physics-rich data, encoding both galaxy properties as well as physical cosmology (e.g. see the recent review in \citealt{Mesinger2019}).  Optimally extracting information from these lightcone images is challenging, since there is no obvious \textit{a priori}, physically-motivated optimal summary statistic.  Since they are non-Gaussian, the common approach of compressing the images into a power spectrum could waste valuable information.

Here we build on previous work  \citep{gillet19, 21cm_ML_Magena20, 21cm_ML_Kwon20, 21cm_ML_LaPlante19, HMV20} training NNs to predict astrophysical parameters directly from 21 cm images.  We introduce recurrent neural networks (RNNs) to this field.  RNNs are designed to characterize temporal evolution by passing through an image sequence with the same set of weights -- allowing them to efficiently find local minima in the loss function (see Appendix \ref{app:layers} for more details). 

To compare the performances of RNNs and traditional CNNs, we construct three datasets with increasing levels of signal contamination (c.f. Figure \ref{fig:observation_pipeline}): (i) {\it Mean Removal}, (ii) {\it + SKA Noise}, (iii) {\it + Horizon Cut}. We train our NN architectures on all three databases using MSE minimization of the parameter predictions.  We vary four astrophysical parameters to generate the underlying cosmic 21 cm signal, capturing the UV and X-ray properties of the first galaxies.  These parameters were chosen as they are the most physically-motivated \enquote{tuning knobs} driving a large variation in possible signals.

We find the RNNs outperform CNNs on images with minimal signal contamination (our {\it Mean Removal} database).  The mean square prediction errors for the best RNN architecture, {\it SummaryRNN} were a factor of $\sim 2$ lower than for a CNN of comparable depth, and up to a factor of $\sim 8$ lower than the previous results applying a shallower CNN on the same database of cosmic signals \citep{gillet19}. It is important to note that these correspond to the prediction errors of point-estimate NNs. In future work we will imploy Bayesian techniques to obtain the full posterior over our parameter space (e.g. \citealt{21cm_ML_Hortua20, zhao2021}).

Using gradient saliency maps, we confirm that our NNs are identifying physically-relevant features when trained.  The networks focus on the correct astrophysical epoch(s) that are relevant for each parameter.

When trained on the signal contaminated images ({\it + SKA Noise} and {\it + Horizon Cut}), parameter prediction becomes less accurate.  However, even in the most pessimistic case, parameters are predicted with reasonable accuracy (with $R^2$ ranging from 0.53--0.97). All architectures perform comparably on the contaminated images, which is likely due to the limited size of our data sets. Moreover, our foreground avoidance technique was fairly conservative; better calibration and foreground removal can improve parameter estimation.  In future work, we will explore these trends further, quantifying the dataset size needed for accurate parameter estimation, given different levels of image contamination.

\section*{Acknowledgements}
This work was supported by the European Research
Council (ERC) under the European Union’s Horizon 2020
research and innovation programme (grant agreement No
638809 – AIDA – PI: Mesinger). The results presented here
reflect the authors’ views; the ERC is not responsible for
their use.  We are grateful to PRACE for awarding us computational resources at Piz Daint, Switzerland, through the PRACE tier-0
grant AIfor21cm (project no. 2019204987). 
We acknowledge the CINECA award under the ISCRA initiative, for the availability of high performance computing resources and support (IsC80\_MGLA21cm, project no. HP10CBAK8I). We also acknowledge computational resources of the HPC center at SNS.

\section*{Data availability}
The data underlying this article will be shared on reasonable request to the corresponding author.

\bibliographystyle{mnras}
\bibliography{reference}

\appendix
\section{Convolutional and LSTM layers} \label{app:layers}

Here we briefly present the general structure of convolutional (Conv), Long Short Term Memory (LSTM) layers, as well as the combination of the two (ConvLSTM).

Let's start from a fully connected (FC) layer, for which we can write a vector transformation as
\begin{equation}
    X_n = \psi ( W_n^{n-1} \cdot X_{n-1} + B_n).
    \label{app:eq:FC}
\end{equation}
Here, $n$ represents a layer index, $X_{n-1}$ and $X_n$ are the input and output vectors, $W_n^{n-1}$ and $B_n$ the weight matrix and the bias vector, and $\psi$ is the activation function.\footnote{For easier notation, we assume that $\psi$ acts element-wise on its input.} 
It is easy to see from Eq. \ref{app:eq:FC} that every neuron takes the full input vector into account, hence the name of a FC layer.

However for an input containing spatially correlated data, it is more efficient to make a transformation which is in some sense local. The use of convolutions is the simplest choice. Each layer is represented by $c^n$ convolutional filters and every \enquote{channel} represents the output of one convolution. For the $i$-th channel $(X_n)_i$ we can write:
\begin{equation}
    (X_n)_i = \psi \left(\sum_{k} (W_n^{n-1})_{ki} * (X_{n-1})_k + (B_n)_i \right),
    \label{app:eq:Conv}
\end{equation}
or in simplified notation: $X_n = \psi(W^n_{n-1} * X^{n-1} + B_n)$. The summation $k$ is performed over $c^{n-1}$ channels. In such way, the output $X_n(\b{d})$ depends locally on $X_{n-1}$ around $\b{d}$, where locality is defined by a convolutional filter. For an intuitive and visual explanation of different convolutional operations, see the review by \cite{dumoulin16}.

Finally, let's consider the case of input $X$, where one of the dimensions (axis) can be considered as time (or in general a sequence). To incorporate that fact directly into layer design, we could compute for example $1\mathrm{D}$ convolutions with respect to the time axis, where correlations are learned on a domain of the actual convolution (for a non-trivial example see \citealt{WaveNet}). However, we could also imagine encoding such information in an {\it iterative} manner. If we start with some \enquote{hidden state} $H_{t-1}$, the operation
\begin{equation}
    H_{t} = \psi (H_{t-1}, X_t; W, B)
    \label{app:eq:RNN}
\end{equation}
would take the input $X_t$ and with already encoded information in $H_{t-1}$, update it into $H_t$. Such operations are the basis of recurrent neural networks (RNN).  

Below we describe the specific choice of recurrent layers we use in our architectures: LSTM and ConvLSTM.  Both are built on the general concepts mentioned above.

\subsection{LSTM} \label{app:LSTM}
\begin{figure}
    \centering
    \includegraphics[width=\linewidth]{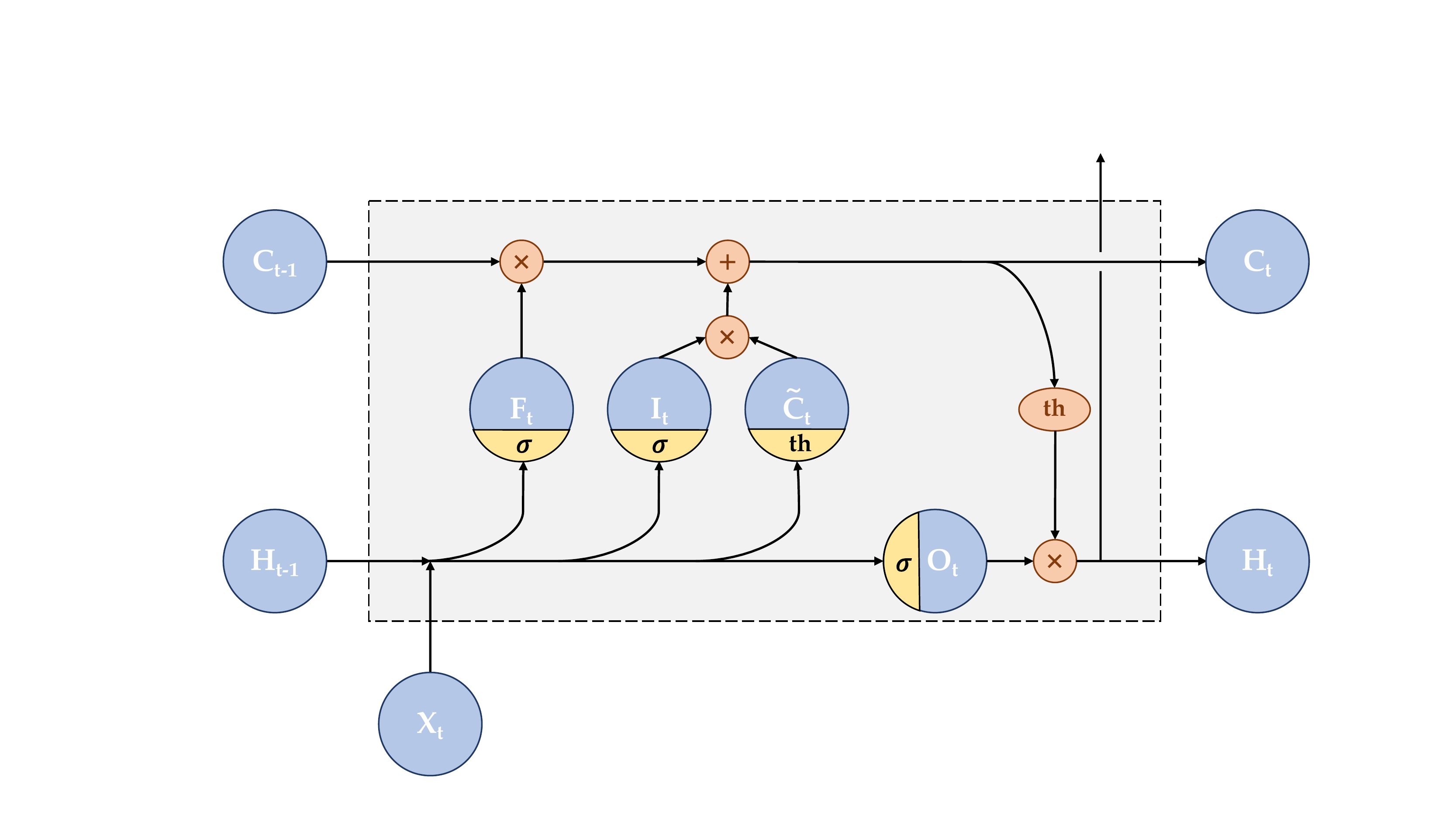}
    \caption{LSTM cell, showing the updating of a cell's states $C_t$ and $H_t$ for one timestep. Upper exiting arrow outputs $H_t$ used for another LSTM layer if needed.}
    \label{fig:LSTM_cell}
\end{figure}

Long Short Term Memory cells \citep{LSTM} solve several problems occurring in simple RNN structures. In particular, the problem of vanishing/exploding gradients is solved by separating short (fast) and long (slow) correlations in the data. In Figure \ref{fig:LSTM_cell} we show a diagram of a LSTM cell, with the following accompanying equations:
\begin{equation}
\begin{aligned}
F_{t} &=\sigma (W^{F}_X \cdot X_{t} + W^F_H \cdot H_{t-1} + B^F) \, , \\
I_{t} &=\sigma(W^{I}_X \cdot X_t + W^I_H \cdot H_{t-1} + B^I )  \, ,\\
\widetilde{C}_{t} &=\operatorname{th} (W^{C}_X \cdot X_t + W^C_H \cdot H_{t-1} + B^{C} )  \, ,\\
O_{t} &=\sigma(W^{O}_X \cdot X_t + W^O_H \cdot H_{t-1}+ B^O)  \, , \\ 
\midrule
C_{t} &= F_{t} \times C_{t-1} + I_{t} \times \widetilde{C}_{t}  \, ,\\
H_{t} &=O_{t} \times \operatorname{th} ( C_{t}) \, ,
\end{aligned}
\end{equation}
where $C_t$, $H_t$ represent the cell's slow and fast hidden states, respectively, and ${W^i_X, W^i_H, B^i}$ are trained weights and biases. The operator $\times$ stands for element-wise product. 
The intuitive motivation behind the \enquote{helping gates} used in LSTM cells are roughly as follows:
\begin{itemize}
    \item[\boldmath$F_t$] - forget gate; determines what parts of state $C$ should be forgotten - for $0$ completely reject previous information, for $1$ leave it untouched,
    \item[\boldmath$I_t$] - input gate; analogous to $F_t$, determines what new information should be added to the state $C$,
    \item[\boldmath$\widetilde{C}_t$] - candidates gate; determines possible new information to be added to  $C$, 
    \item[\boldmath$O_t$] - output gate; determines what values of the updated state $C$ should be passed to the state $H$.
\end{itemize}
Interested readers can refer to \cite{jozefowicz2015empirical} for an investigation of alternate designs.

Finally, we can stack multiple LSTMs to \enquote{deepen} the network by remembering fast hidden states $H_t$ for the whole sequence and passing it as an input to the next LSTM cell (the connection is denoted by the upward arrow in Figure \ref{fig:LSTM_cell}). 

\subsection{ConvLSTM} \label{app:ConvLSTM}
In the case of temporal \textit{and} spatially correlated data, we might require convolutional LSTM layers \citep{ConvLSTM}. Convolutional LSTM layers have an identical structure to the generic LSTM discussed above, with the change of weight matrices representing convolutions over spatial dimensions. We can therefore write:
\begin{equation}
\begin{aligned}
F_{t} &=\sigma (W^{F}_X * X_{t} + W^F_H * H_{t-1} + B^F) \, , \\
I_{t} &=\sigma(W^{I}_X * X_t + W^I_H * H_{t-1} + B^I )  \, ,\\
\widetilde{C}_{t} &=\operatorname{th} (W^{C}_X * X_t + W^C_H * H_{t-1} + B^{C} )  \, ,\\
O_{t} &=\sigma(W^{O}_X * X_t + W^O_H * H_{t-1}+ B^O)  \, .
\end{aligned}
\end{equation}
Updates of the cell's states follow same equations as before. We note that in this research we are not using \enquote{peephole connections} (e.g. \citealt{ConvLSTM}).

\section{Detailed Architectures} \label{app:DetailedArchitectures}
Here we present the detailed structure of all of the three NNs we use in this work. 
CNN, ConvRNN and SummaryRNN architectures are summarized in tables \ref{table:layers_CNN}, \ref{table:layers_ConvRNN}, \ref{table:layers_SummaryRNN}, respectively. 
\enquote{Layer Shape} describes a kernel size (for Conv, MaxPool and ConvLSTM layers) or a number of neurons (for LSTM and FC layers).
\enquote{Tensor Shape} denotes tensor dimensions after a particular layer. The input tensor consists of frequency, two spatial sky-plane components and \enquote{channels} dimensions, respectively.

\bsp
\newpage
\begin{table}
\centering
\begin{tabular}{r|c|c|r} 
    Layer Type & Layer Shape & Tensor Shape & Params \\ 
    \hline\hline
Input &  & $(526, 25, 25, 1)$ & $0$         \\
\hline
$3\mathrm{D}$ Conv & $(8, 8, 8)$ & $(519, 18, 18, 128)$ &  $65~664$      \\
\hline
$3\mathrm{D}$ MaxPool  & $(2, 2, 4)$ & $(129, 9, 9, 128)$  &   $0$         \\
\hline
Batch Norm & & $(129, 9, 9, 128)$ &     $256$       \\
\hline
$2\mathrm{D}$ Conv & $(4, 4, 1)$ & $(129, 6, 6, 128)$ &  $262~272$      \\
\hline
$2\mathrm{D}$ MaxPool  & $(2, 2, 1)$ & $(129, 3, 3, 128)$  &   $0$         \\
\hline
Batch Norm & & $(129, 3, 3, 128)$ &     $256$       \\
\hline
$1\mathrm{D}$ Conv & $(1, 1, 4)$ & $(126, 3, 3, 128)$ &  $65~664$      \\
\hline
$1\mathrm{D}$ MaxPool  & $(1, 1, 2)$ & $(63, 3, 3, 128)$  &   $0$         \\
\hline
Batch Norm & & $(63, 3, 3, 128)$ &     $256$       \\
\hline
$2\mathrm{D}$ Conv & $(3, 3, 1)$ & $(63, 1, 1, 128)$ &  $147~584$      \\
\hline
Batch Norm & & $(63, 1, 1, 128)$ &     $256$       \\
\hline
$1\mathrm{D}$ Conv & $(1, 1, 4)$ & $(60, 1, 1, 128)$ &  $65~664$      \\
\hline
$1\mathrm{D}$ MaxPool  & $(1, 1, 2)$ & $(30, 1, 1, 128)$  &   $0$         \\
\hline
Batch Norm & & $(30, 1, 1, 128)$ &     $256$       \\
\hline
$1\mathrm{D}$ Conv & $(1, 1, 4)$ & $(27, 1, 1, 128)$ &  $65~664$      \\
\hline
$1\mathrm{D}$ MaxPool  & $(1, 1, 2)$ & $(13, 1, 1, 128)$  &   $0$         \\
\hline
Batch Norm & & $(13, 1, 1, 128)$ &     $256$       \\
\hline
$1\mathrm{D}$ Conv & $(1, 1, 4)$ & $(10, 1, 1, 128)$ &  $65~664$      \\
\hline
$1\mathrm{D}$ MaxPool  & $(1, 1, 2)$ & $(5, 1, 1, 128)$  &   $0$         \\
\hline
Batch Norm & & $(5, 1, 1, 128)$ &     $256$       \\
\hline
Flatten  &  & $(640)$ &        $0$         \\
\hline
Dropout  &  & $(640)$   &      $0$         \\
\hline
FC & $512$ & $(512)$  &              $328~192$      \\
\hline
Batch Norm & & $(512)$ &               $1~024$       \\
\hline
FC & $256$ & $(256)$  &              $131~328$       \\
\hline
Batch Norm & & $(256)$ &               $512$       \\
\hline
FC & $64$ & $(64)$  &               $16~448$       \\
\hline
FC & $8$ & $(16)$  &               $1~040$       \\
\hline
FC & $4$ & $(4)$   &              $68$        \\
\hline\hline
\multicolumn{3}{r}{Total:} & $1~218~580$ \\
\end{tabular}
\caption{Best CNN model. The input tensor consists of frequency, two spatial sky-plane components and \enquote{channels} dimensions, respectively. }
\label{table:layers_CNN}
\end{table}

\begin{table}
\centering
\begin{tabular}{r|c|c|r} 
    Layer Type & Layer Shape & Tensor Shape & Params \\ 
    \hline\hline
Input &  & $(526, 25, 25, 1)$ & $0$         \\
\hline
$2\mathrm{D}$ ConvLSTM  & $(8, 8)$ & $(526, 18, 18, 32)$ &  $270~464$      \\
\hline
$2\mathrm{D}$ MaxPool (TD) & $(2, 2)$ & $(526, 9, 9, 32)$  &   $0$         \\
\hline
Batch Norm & & $(526, 9, 9, 32)$ &     $64$       \\
\hline
$2\mathrm{D}$ ConvLSTM &  $(4, 4)$ &    $(526, 6, 6, 64)$ &    $393~472$    \\
\hline
$2\mathrm{D}$ MaxPool (TD)  &  $(2, 2)$ &$(526, 3, 3, 64)$  &  $0$         \\
\hline
Batch Norm &  & $(526, 3, 3, 64)$  &  $128$       \\
\hline
Flatten (TD)  &  & $(526, 576)$ &        $0$         \\
\hline
Dropout (TD)  &  & $(526, 1576)$   &      $0$         \\
\hline
LSTM & $128$ &      $(526, 128)$ &          $361~472$    \\
\hline
Batch Norm & & $(526, 128)$ &         $256$       \\
\hline
LSTM & $128$ &      $(526, 128)$ &           $132~096$     \\
\hline
Batch Norm & & $(526, 128)$ &           $256$       \\
\hline
LSTM & $64$ &      $(526, 64)$  &         $49~664$     \\
\hline
Batch Norm & & $(526, 64)$ &           $128$       \\
\hline
LSTM & $64$ &      $(526, 64)$  &         $33~280$     \\
\hline
Batch Norm & & $(526, 64)$ &           $128$       \\
\hline
LSTM\tablefootnote{\label{note1} Keeping only the final hidden state at the end.} & $32$ &      $(32)$ &                $12~544$     \\
\hline
Batch Norm & & $(32)$ &               $64$       \\
\hline
FC & $32$ & $(32)$  &              $1~056$      \\
\hline
FC & $16$ & $(16)$  &              $528$       \\
\hline
FC & $8$ & $(8)$  &               $136$       \\
\hline
FC & $4$ & $(4)$   &              $36$        \\
\hline\hline
\multicolumn{3}{r}{Total:} & $1~255~772$ \\
\end{tabular}
\caption{Best ConvRNN model. The input tensor consists of frequency, two spatial sky-plane components and \enquote{channels} dimensions, respectively. \enquote{TD} labels time distributed layer, meaning it is shared between frequencies.}
\label{table:layers_ConvRNN}
\end{table}

\begin{table}
\centering
\begin{tabular}{r|c|c|r} 
    Layer Type & Layer Shape & Tensor Shape & Params \\ 
    \hline\hline
Input &  & $(526, 25, 25, 1)$ & $0$         \\
\hline
$2\mathrm{D}$ Conv (TD)  & $(8, 8)$ & $(526, 18, 18, 64)$ &  $4~160$      \\
\hline
$2\mathrm{D}$ MaxPool (TD)  & $(2, 2)$ & $(526, 9, 9, 64)$  &   $0$         \\
\hline
Batch Norm & & $(526, 9, 9, 64)$ &     $128$       \\
\hline
$2\mathrm{D}$ Conv (TD) &  $(4, 4)$ &    $(526, 6, 6, 128)$ &    $131~200$    \\
\hline
$2\mathrm{D}$ MaxPool (TD)  &  $(2, 2)$ &$(526, 3, 3, 128)$  &  $0$         \\
\hline
Batch Norm &  & $(526, 3, 3, 128)$  &  $256$       \\
\hline
Flatten (TD)  &  & $(526, 1152)$ &        $0$         \\
\hline
Dropout (TD)  &  & $(526, 1152)$   &      $0$         \\
\hline
FC (TD)\tablefootnote{Final layer of the $2\mathrm{D}$ compression, i.e. summary space.} & $128$ & $(526, 128)$ &         $147~584$    \\
\hline
Batch Norm & & $(526, 128)$ &          $256$       \\
\hline
LSTM & $128$ &      $(526, 128)$ &          $132~096$    \\
\hline
Batch Norm & & $(526, 128)$ &         $256$       \\
\hline
LSTM & $128$ &      $(526, 64)$ &           $49~664$     \\
\hline
Batch Norm & & $(526, 64)$ &           $128$       \\
\hline
LSTM & $64$ &      $(526, 64)$  &         $33~280$     \\
\hline
Batch Norm & & $(526, 64)$ &           $128$       \\
\hline
LSTM\textsuperscript{\ref{note1}} & $32$ &      $(32)$ &                $12~544$     \\
\hline
Batch Norm & & $(32)$ &               $64$       \\
\hline
FC & $32$ & $(32)$  &              $1~056$      \\
\hline
FC & $16$ & $(16)$  &              $528$       \\
\hline
FC & $8$ & $(8)$  &               $136$       \\
\hline
FC & $4$ & $(4)$   &              $36$        \\
\hline\hline
\multicolumn{3}{r}{Total:} & $513~500$ \\
\end{tabular}
\caption{Best SummaryRNN model. The input tensor consists of frequency, two spatial sky-plane components and \enquote{channels} dimensions, respectively. \enquote{TD} labels time distributed layer.}
\label{table:layers_SummaryRNN}
\end{table}

\label{lastpage}
\end{document}